\documentclass[3p, 12pt, authoryear]{elsarticle}

\usepackage{amsthm}
\usepackage{amsmath}
\usepackage{amssymb}
\usepackage{graphicx}

\usepackage{enumerate}
\usepackage{enumitem}
\usepackage{lscape}
\usepackage{rotating}
\usepackage{booktabs}
\usepackage{multirow}
\usepackage{makecell}
\usepackage{subfigure}
\usepackage{tikz}
\usepackage{arydshln}

\usetikzlibrary{positioning,shapes,calc}
\usepackage{fix-cm}
\usepackage{epsf}
\usepackage{psfrag}
\usepackage{pstool}
\usepackage{pifont}

\usepackage{epstopdf}
\usepackage{epsfig}

\usepackage[ruled,vlined]{algorithm2e}
\usepackage[hyperindex,breaklinks,colorlinks,citecolor=blue,urlcolor=black]{hyperref}

\usepackage{listings}

\usepackage{datatool}
\usepackage{natbib}

\DTLnewdb{ref}

\def\bibnote#1#2{%
  \DTLnewrow{ref}
  \DTLnewdbentry{ref}{mylabel}{#1}
  \DTLnewdbentry{ref}{mynote}{#2}
}

\makeatletter
\patchcmd{\@lbibitem}%
  {\item[\hfil\NAT@anchor{#2}{\NAT@num}]}%
  {%
    \item[\hfil\NAT@anchor{#2}{\NAT@num}]%
    \DTLforeach[\DTLiseq{\mylabel}{#2}]{ref}{\mylabel=mylabel,\mynote=mynote}{\mynote}
  }{}{\message{Failed to insert prefix to reference}}%

\SetKwInput{KwData}{Initialize}

\newcommand{\ind}{\stackrel{\rm ind}{\sim}}
\newcommand{\iid}{\stackrel{\rm iid}{\sim}}

\newcommand{\x}{\mathbf{x}}

\newcommand{\w}{\mathbf{w}}

\newcommand{\z}{\mathbf{z}}

\newcommand{\C}{\mathbf{C}}
\newcommand{\V}{\mathbf{V}}
\newcommand{\Li}{\mathbf{L}}
\newcommand{\bal}{\boldsymbol{\alpha}}
\newcommand{\bbe}{\boldsymbol{\beta}}

\newcommand{\bga}{\boldsymbol{\gamma}}
\newcommand{\beps}{\boldsymbol{\epsilon}}

\newcommand{\bPsi}{\boldsymbol{\Psi}}

\newcommand{\btau}{\boldsymbol{\tau}}
\newcommand{\bphi}{\boldsymbol{\phi}}
\newcommand{\bgam}{\boldsymbol{\gamma}}

\newcommand{\bom}{\boldsymbol{\omega}}

\newcommand{\A}{\mathbf{A}}
\newcommand{\B}{\mathbf{B}}
\renewcommand{\b}{\mathbf{b}}

\newcommand{\T}{\top}
\newcommand{\bT}{\mathbf{T}}
\newcommand{\D}{\mathbf{D}}
\newcommand{\R}{\mathbf{R}}
\newcommand{\bR}{\mathbb{R}}
\newcommand{\PP}{\mathbf{P}}
\newcommand{\W}{\mathbf{W}}
\newcommand{\X}{\mathbf{X}}
\newcommand{\Y}{\mathbf{Y}}
\newcommand{\U}{\mathbf{U}}
\newcommand{\Z}{\mathbf{Z}}
\newcommand{\I}{\mathbf{I}}

\newcommand{\bs}{\boldsymbol}
\newcommand{\mf}{\mathbf}
\renewcommand{\*}{^\star}

\journal{Computational Statistics \& Data Analysis}

\begin{document}
	
	\begin{frontmatter}
		
		
		\title{Functional clustering methods for binary longitudinal data with temporal heterogeneity}
		
		\author[add1]{Jinwon Sohn}
		\ead{sohn24@purdue.edu}
		\author[add2,add3]{Seonghyun Jeong\corref{cor1}}
		\ead{sjeong@yonsei.ac.kr}
		\author[add4]{Young Min Cho}
		\ead{jch0@seas.upenn.edu}
		\author[add2,add3]{Taeyoung Park\corref{cor2}}
		\ead{tpark@yonsei.ac.kr}
		
		\cortext[cor1]{Equally contributed with the first author.}	
		\cortext[cor2]{Corresponding author}

		\address[add1]{Department of Statistics, Purdue University, IN 47907, USA}
		\address[add2]{Department of Applied Statistics, Yonsei University, Seoul 03722, Korea}
		\address[add3]{Department of Statistics and Data Science, Yonsei University, Seoul 03722, Korea}
		\address[add4]{Department of Computer and Information Science, University of Pennsylvania, PA 19104, USA}


		\begin{abstract}
In the analysis of binary longitudinal data, it is of interest to model a dynamic relationship between a response and covariates as a function of time, while also investigating similar patterns of time-dependent interactions. We present a novel generalized varying-coefficient model that accounts for within-subject variability and simultaneously clusters varying-coefficient functions, without restricting the number of clusters nor overfitting the data. In the analysis of a heterogeneous series of binary data, the model extracts population-level fixed effects, cluster-level varying effects, and subject-level random effects. Various simulation studies show the validity and utility of the proposed method to correctly specify cluster-specific varying-coefficients when the number of clusters is unknown. The proposed method is applied to a heterogeneous series of binary data in the German Socioeconomic Panel (GSOEP) study, where we identify three major clusters demonstrating the different varying effects of socioeconomic predictors as a function of age on the working status. 
		\end{abstract}
		
		
	\end{frontmatter}

	\section{Introduction}
	\label{sec:intro}
	
	Mixed-effects models are commonly used in binary longitudinal studies in the social, behavioral, and health sciences. These models' popularity stems from their ability to capture longitudinal effects generated by repeated-measurement processes. To be more specific, random effects are introduced into linear models with only fixed effects to reflect the correlation between observations on the same subject. This extension also avoids some of the technical issues that can arise during the analysis of variance. For example, \cite{stir:etal:84} showed that mixed-effects models have advantages over Markov models when dealing with a series of binary data because they are better at interpreting the effects of covariates and circumventing some of the difficult issues caused by unbalanced design or missing values. Other approaches to dealing with serial effects in longitudinal data provide practical recommendations by combining the mixed-effects models with other statistical methods \citep{vari:czad:09,guer:etal:12}.
	
	A varying-coefficient model has been shown to be extremely effective for modeling of time-varying effects in longitudinal studies \citep{hast:tibs:93, hoov:etal:98, wu:etal:98, lang:brez:04, yanq:huli:05, fan:zhan:08, lu:zhan:09, jeon:park:16, jeon:etal:17,park:jeon:17}.
Such varying-coefficient functions can be easily modeled by Bayesian methods, e.g., Bayesian P-splines \citep{lang:brez:04}, series priors \citep{shen:ghos:15}, Gaussian process priors \citep{neal:98}, Bayesian wavelets \citep{chip:etal:97}, and free-knot splines and adaptive knot selection \citep{smit:kohn:96,dima:etal:01}. The main advantage of using the Bayesian approaches is that uncertainty quantification is naturally performed with credible sets obtained by Markov chain Monte Carlo (MCMC). The method of free-knot splines and adaptive knot selection, in particular, exhibits natural local adaptation to spatially inhomogeneous smoothness \citep{smit:kohn:96,kohn:etal:01,rupp:etal:03,kang:jeon:23}.

	Traditional varying-coefficient mixed models focus on exploring common varying coefficients shared across all subjects. However, because there are often various sources of heterogeneity among subjects, particularly in longitudinal studies, such a common structure of varying coefficients may be oversimplified, leading to incorrect conclusions. To uncover the heterogeneity of the population, many model-based clustering approaches have been proposed from both frequentist and Bayesian perspectives \citep{lenk:desa:00, jame:suga:03, hear:etal:06, shi:wang:08, assm:jens:11, coff:etal:14, berr:etal:22}. For example, \cite{jame:suga:03} suggested modeling individual basis coefficients by random effects with the mean indexed by the cluster. \cite{hear:etal:06} developed a hierarchical Bayesian model that avoids MCMC using their particular model formulation. \cite{coff:etal:14} developed a clustering method for longitudinal gene profiles via penalized splines. More recently, \cite{berr:etal:22} devised a semi-parametric mixture model with mixture weights and conditional means that are modeled as nonlinear functions of covariates. Although these frameworks clearly offer inferential advantages in the presence of heterogeneity, most require determining the appropriate number of clusters and  additional evaluation steps decoupled from parameter estimation such as cross validation. 
	
In contrast, the Bayesian nonparametric framework naturally chooses the required number of clusters in a data-driven way by using stochastic process priors that randomly partition a sample space to be clustered. These priors include the Dirichlet process (DP) \citep{ferg:73}, the two-parameter Poisson-Dirichlet process \citep{pitm:yor:97}, and the generalized stick-breaking process \citep{ishw:jame:01}, to name a few. As they are basically infinite-dimensional priors, they have become essential clustering tools for modeling an infinite number of clusters in various areas \citep{ishw:jame:01, teh:etal:06,  lau:gree:07, wall:etal:10, cana:duns:11, yere:etal:14,kyun:15}. In the context of nonparametric regression, \cite{mull:etal:96} employed the DP prior to jointly partition the support of response and predictor variables, which performs locally weighted regression estimation in terms of Bayesian predictive inference. \cite{gelf:etal:05} incorporates the dependent DP prior \citep{mace:etal:01} into the Gaussian process prior for spatial analysis. \cite{ray:mall:06} studies the Bayesian wavelet regression model where the DP prior has a base measure that expedites sparsity of the wavelet coefficients. \cite{petr:etal:09} further considered the local heterogeneity in a subgroup of curves by proposing a hybrid Dirichlet prior that overcomes the global heterogeneity. \cite{chib:green:10} used the DP prior for modeling an error distribution while approximating nonlinear components via cubic splines with a smoothness prior regularizing difference of spline coefficients at the knot locations. \cite{rodr:duns:14} employed the generalized DP prior to cluster curves smoothed by the free-knot spline method. \cite{suar:ghos:16} assigned the DP prior on each wavelet  coefficient independently not jointly on the set of coefficients with the sparsity structure used in \cite{ray:mall:06}. \cite{marg:etal:21} applied the DP prior to the clustering of functional principle scores, which improves the curve and correlation reconstruction.

While functional clustering for continuous response has emerged, there have been relatively fewer works for longitudinal binary responses with nonparametric components. \cite{kuss:etal:06} considered a parametric logistic model where the cluster allocation is assumed to follow a multinomial distribution. Similarly, \cite{assm:jens:11} designed a Bayesian probit regression model with the multinomial label allocation. On the one hand, \cite{hann:etal:11} proposed a DP mixture model for generalized linear models in the spirit of \cite{mull:etal:96}, with the restriction of a linear relationship within each cluster. More recently, \cite{zhu:etal:21} devised a model-free clustering method for binary longitudinal data using a pairwise penalty to nearby clusters, but their model does not account for any functional effects. We find that none of the aforementioned studies deal with both model-based clustering on a mixed-effects model (especially with the DP prior) and nonparametric function estimation with guaranteed smoothness.

Our contribution is threefold. 
First, we propose a flexible framework for simultaneously modeling population-level fixed effects, cluster-level varying effects, and subject-level random effects in the analysis of binary longitudinal data. The proposed model is a probit varying-coefficient mixed model that flexibly and adaptively identifies different subpopulations having their own varying-coefficient functions that can be either constant, linear, or nonlinear. 
Second, we devise new prior distributions for the posterior analysis and effective functional clustering of the proposed model. In particular, it is well known that the measurement scale of data must be considered in choosing the base measure for the DP prior \citep{gelm:etal:15}. We carefully design a prior distribution with a reasonable scale so that it can start a new cluster well within a sampler to account for an infinite number of clusters, while achieving suitable smoothing for function estimation with spatial adaptation.
Third, we construct a partially collapsed Gibbs (PCG) sampler to cover the varying-dimensional parameter issue of a standard Gibbs sampler and facilitate posterior computation via the method of partial collapse \citep{vand:park:08, park:vand:09}. To maintain a transition kernel, a PCG sampler, unlike a standard Gibbs sampler, requires its steps to be performed in a specific order. We thus develop a PCG sampler that can be used in the fitting of the proposed model.


	
	The remainder of this paper is organized as follows. In Section~\ref{sec:model}, we describe the probit varying-coefficient mixed model and discuss how the DP prior constructs model-based clustering. Section~\ref{sec:bayes} specifies prior distributions and constructs efficient sampling steps based on the method of partial collapse. In Section~\ref{sec:sim}, simulation studies are presented to validate the proposed method. Section~\ref{sec:app} applies the proposed model to the GSOEP data, and Section~\ref{sec:disc} discusses the results. \ref{sec:appen} contains a detailed description of the proposed method, while \ref{sec:appen_rpackage}  describes how to install the R package for the proposed method. The R package is currently available on the first author's github\footnote{https://github.com/Jwsohn612/fvcc}  to demonstrate that all of the results in Sections~\ref{sec:sim} and~\ref{sec:app} can be reproduced.

	\section{Probit Varying-Coefficient Mixed Models for Functional Clustering}
	\label{sec:model}
	
	Let $Y_{ij}$ represent a binary response observed at time $t_{ij}$ for observation $j$ on subject $i$, where $i=1,\dots,N$ and $j=1,\dots,n_i$. The outcome of the response $Y_{ij}$ can be expressed as an indicator function of the sign of a latent variable $L_{ij}$, i.e.,
	\begin{align*}
		Y_{ij} ~=~ I(L_{ij} > 0),
	\end{align*}
	where the latent variable is introduced for computational convenience but can be interpreted as a utility difference between choosing $Y_{ij}=1$ or 0.
	
	In longitudinal studies, a relationship between the latent variable $L_{ij}$ and the available covariates is commonly specified by a generalized linear mixed model to account for between-subject variability. Specifically, the generalized linear mixed model is expressed as
	\begin{align}
		L_{ij} ~=~ \X_{i}^{(j)}\bbe + \Z_{i}^{(j)}\b_{i} + \epsilon_{ij},
		\label{eq:mm}
	\end{align}
	where $\X_{i}=(\x_{i1}, \dots, \x_{iq})$ and $\Z_{i}=(\z_{i1}, \dots, \z_{ir})$ are $n_{i} \times q,$ and $n_{i}\times r$ design matrices for subject $i$,  $\X_{i}^{(j)}$ and $\Z_{i}^{(j)}$ denote the $j$th row vectors of $\X_i$ and $\Z_{i}$, respectively, $\bbe$ is a $q\times1$ vector of fixed effects, $\b_{i}$ is a $r\times1$ vector of random effects for subject $i$, and $\epsilon_{ij}$ is an underlying error term that is assumed to follow a logistic or normal distribution in the logit or probit model, respectively. Note that the model in~(\ref{eq:mm}) involves multiple random effects on binary longitudinal data; for linear mixed models with multiple random effects, see \cite{vine:etal:96,meng:vand:98, vand:00, kim:etal:13, park:min:16}.
	
	In the presence of within-subject correlation over time, \citet{jeon:etal:17} extends the generalized linear mixed model in~(\ref{eq:mm}) to incorporate varying-coefficients $\bal(t)$ that vary over time $t$, i.e.,
	\begin{align}
		L_{ij}
		~=~ \W_{i}^{(j)}\bal(t_{ij}) + \X_{i}^{(j)}\bbe +\Z_{i}^{(j)}\b_{i} + \epsilon_{ij},
		\label{eq:vcmm}
	\end{align}
	where $\W_{i}=(\w_{i1}, \dots, \w_{ip})$ is $n_{i} \times p$ design matrices for subject $i$, $\W_{i}^{(j)}$ denotes the $j$th row vector of $\W_i$, $\bal(t)=(\alpha_{1}(t),\dots, \alpha_{p}(t))^{\T}$ is a $p\times1$ vector of unknown smooth functions that vary over time $t$, and $t_{ij}$ is the $j$th time of the $i$th subject. Note that $\bal(t_{ij})$ is a vector of real values of $\bal(t)$ evaluated at $t_{ij}$. The $l$th time-varying function in the vector $\bal(t)$, i.e., $\bal_{l}(t)$ can be modeled with regression splines that use a linear combination of basis functions, e.g.,
	\begin{align}
		\B_{l}(t) ~=~ \left( 1, t, \left|t-\omega_{l1}\right|^3, \left|t-\omega_{l2}\right|^3,\dots,\left|t-\omega_{lM_{l}}\right|^3 \right)^\top,
		\label{eq:basis}
	\end{align}
	where $\bom_l=(\omega_{l1},\dots,\omega_{lM_{l}})$ is an ordered sequence of knot-candidate locations within the range of observed time points, for $l=1,\dots,p$. The amount of smoothness for the $\l$th regression spline is controlled by the number $M_l$ and locations $\bom_l$. 
	
	To account for heterogeneity among the subjects while borrowing strength across the different subjects, we consider allocating each subject to its own cluster with different functions of varying-coefficients. To do so, we represent the set of functions $\bal(t)$ in~(\ref{eq:vcmm}) as the subject-level varying-coefficients, i.e., $\bal_i(t)=(\alpha_{i1}(t),\dots, \alpha_{ip}(t))^{\T}$.	Each of the unknown subject-specific varying-coefficient functions is assumed to fall in the linear span of a set of its own basis functions according to basis selection, i.e.,
	\begin{align*}
\alpha_{il}(t) ~\approx~ (\B_{l}(t)\odot\bgam_{il})^\top \bphi_{il},
	\end{align*}
where $\odot$ denotes element-wise multiplication of vectors in accordance with values of $\bgam_{il}$, $\bgam_{il}=(1,\gamma_{il0},\gamma_{il1},\dots,\gamma_{ilM_l})^\top$ denotes an $(M_l+2)\times 1$ vector of indicator variables for basis inclusion, $\gamma_{ilm}=1$ represents that the $(m+2)$th element in $\bgam_{il}$ is used as a basis function, and $\bphi_{il}$ denotes an $(M_l+2)\times 1$ vector of basis coefficients corresponding to the $l$th varying-coefficient for subject $i$. The first element in $\bgam_{il}$ equals one, so the constant basis function in (\ref{eq:basis}) always remains in the model. When the corresponding covariate has no interaction with time,  we have $\gamma_{ilm}=0$ for $m=0,\dots,M_l$, and 
the model in~(\ref{eq:vcmm}) is reduced to the generalized linear mixed model in~(\ref{eq:mm}). That is, when the true varying-coefficient function is constant, our model can estimate it as a constant function by choosing $\gamma_{ilm}=0$ for $m=0,\dots,M_l$, reducing modeling bias and avoiding the possibility of overfitting. When the true varying-coefficient function is nonlinear, selecting appropriate knots allows the estimated function to adapt to the true one's curvature. That is, we use data to adjust the spatially inhomogeneous smoothness of a varying-coefficient function, so that more knots are used in a high-curvature region and fewer knots in a low-curvature region. 
This implies that we do not need to a priori determine whether a varying-coefficient function is constant, linear, or nonlinear \citep{jeon:park:16, jeon:etal:17,jeon:etal:21}.
The value of $M_l$ for the knot-candidates is not crucial as long as it is large enough to capture the global and local characteristics of a function. Following the literature \citep[e.g.,][]{kohn:etal:01}, we recommend using 20 to 30 knot-candidates chosen by the sample quantiles of the time variable $t$. If the time variable is repeatedly observed at some points in time, $M_l$ should not be larger than the number of the non-duplicated values for $t$.

	Next, the individualized vector of functions $\bal_{i}(t)$ is given the DP prior, which induces functional clustering with respect to the functions. Section~\ref{sec:2.3} describes how the model leverages the DP prior to cluster varying-coefficients in detail. Let $C_i=k$ denote that subject $i$ belongs to cluster $k$ sharing identical basis functions for varying-coefficients, for $i=1,\dots,N$ and $k=1,\dots,K$. Then through the DP prior, we have $\bal_i(t)=\bal^*_{k}(t)$, which implies $\bgam_{il} = \bgam^*_{kl}$ and $\bphi_{il}=\bphi^*_{kl}$ as well. To be specific,
	\begin{align}
		\alpha_{il}(t) ~=~ \alpha^*_{kl}(t)  ~\approx~ (\B_{l}(t)\odot\bgam^*_{kl})^\top \bphi^*_{kl},
		\quad l=1,\dots,p,
		\label{eq:spline2}
	\end{align}
	for the $i$th subject who is allocated to cluster $k$, having $C_i=k$.
	Thus, for the $i$th subject, the model in~(\ref{eq:vcmm}) can be represented in a matrix form,
	\begin{align}
		\Li_{i} ~=~ \sum_{l=1}^{p}\left(\w_{il} \odot \alpha^*_{C_{i}l}(t_{ij})|_{j=1}^{n_i}\right) + \X_{i}\bbe + \Z_{i}\b_{i} + \beps_{i},
				\label{eq:vcmm1}
	\end{align}	
where $\alpha^*_{C_{i}l}(t_{ij})|_{j=1}^{n_i}=\left(\alpha^*_{C_{i}l}(t_{i1}),\dots, \alpha^*_{C_{i}l}(t_{in_i})\right)^{\top}$ is a $n_i \times 1$ vector of real values,  $\Li_{i}$ is an $n_{i}\times1$ vector of latent variables, and $\beps_{i}$ is a $n_{i}\times1$ error vector. This representation implies that the clustering process is implemented with information about only dynamic covariates.
 
To express \eqref{eq:vcmm1} with the approximation in~\eqref{eq:spline2}, we define an $n_i \times (M_l+2)$ matrix $\B^{*}_{il}=\left(\oplus_{j=1}^{n_i}(\B_{l}(t_{ij})\odot\bgam^*_{C_i l})\right)^{\top}$ where $\oplus$ represents the direct sum of vectors or matrices, for the $l$th covariate of the $i$th subject. It is obvious that this matrix can have zero column vectors when the corresponding elements of $\bgam^*_{C_i l}$ are zero. By removing the columns of 0's, we can obtain an $n_i \times |\bgam^*_{C_i l}|$ submatrix of $\B^{*}_{il}$, which is denoted by $\B^{\star}_{il}$, where $|\bga_{C_i l}^*|=\sum_{m}\gamma^*_{C_{i}lm}$.
	Then the model in~\eqref{eq:vcmm1} can be written as
	\begin{align}
		\Li_{i} ~=~  \W_{i(\bga_{C_i}^*)}^{\star}\bphi_{(\bga_{C_i}^{*})}^{\star} + \X_{i}\bbe + \Z_{i}\b_{i} + \boldsymbol{\epsilon}_{i},
		\label{eq:vcmm2}
	\end{align}
	where $\bphi_{(\bga_{C_i}^{*})}^{\star}$ is a vector of cluster-level basis coefficients whose size is the sum of all elements of $\bga_{C_i}^{*}$. It is
	\begin{align*}
		\bphi_{(\bga_{C_i}^{*})}^{\star} ~=~ \left(\bphi^*_{\bgam^*_{C_{i} 1}} ,\dots ,\bphi^*_{\bgam^*_{C_{i} p}} \right)^\top  ~\in~ \bR^{|\bga_{C_i}^*| \times 1},
	\end{align*}
	where $|\bga_{C_i}^*|=\sum_{l,m}\gamma^*_{C_{i}lm}$ and $\bphi^*_{\bgam^*_{C_{i} l}}$ is a $|\bga_{C_i l}^*|\times 1$ subvector of $\bphi^*_{C_i l}$ whose elements correspond to nonzero columns of $\B^{*}_{il}$. 
	Then, the design matrix $\W_{i(\bga_{C_i}^*)}^{\star}$ is constructed by multiplying each set of selected basis terms to each column of $\W_{i}$, i.e.,
	\begin{align*}
		\W_{i(\bga_{C_i}^*)}^{\star} ~=~ 
		\left[\w_{i1}{\bf 1}_{|\bgam^*_{C_{i} 1}|}^{\top} \odot \B_{i1}^{\star}, \dots, \w_{ip}{\bf 1}_{|\bgam^*_{C_{i} p}|}^{\top} \odot\B_{ip}^{\star} \right]~\in~ \bR^{n_{i}\times |\bga_{C_i}^*|}.
	\end{align*}
where ${\bf 1}_{|\bgam^*_{C_{i} l}|}$ is a vector of ones in ${\mathbb R}^{|\bgam^*_{C_{i} l}| \times 1}$.


	\section{Bayesian Analysis}
	\label{sec:bayes}

	\subsection{Dirichlet Process Prior}
	\label{sec:2.3}
	
	In this paper, the DP is used as a prior distribution to cluster $\bal_{i}(t)$, and this clustering procedure is equivalent to clustering the set of $(\bphi_{i}, \bgam_{i})$, where $\bphi_{i}=(\bphi_{i 1},\dots,  \bphi_{i p})^{\top}$ and  $\bga_{i}=(\bga_{i 1},\dots,  \bga_{i p})^{\top}$, as implied in~(\ref{eq:spline2}). This process assigning the DP prior to the set of $(\bphi_{i}, \bgam_{i})$ is expressed as
	\begin{align*}
		(\bphi_{i}, \bga_{i})|\mathcal H &~\iid~ \mathcal H, \quad i=1,\dots,N,\\
		\mathcal H &~\sim~ {\rm DP} (\nu, \mathcal H_{0}),
	\end{align*}
	where $\nu > 0$ and $\mathcal H_{0}$ is a base distribution that randomly generates cluster-level parameters for $(\bphi_{i}, \bga_{i})$. As another representation of the DP, it is worthwhile to look at the stick-breaking process \citep{seth:94, ishw:jame:01} that allows the truncation of the summation in the DP after a large $K$ component, i.e.,
	\begin{align}					
		\mathcal H(\cdot) ~=~ \sum_{k=1}^{\infty}\pi_{k}\delta_{(\bphi_{k}^{*}, \bga_{k}^{*})}(\cdot)	~\approx~ \sum_{k=1}^{K}\pi_{k}\delta_{(\bphi_{k}^{*}, \bga_{k}^{*})}(\cdot), \quad 	(\bphi_{k}^{*}, \bga_{k}^{*})~\iid~ \mathcal H_{0},\label{eq:pyp1}
	\end{align}	
	where $\bphi_{k}^*=(\bphi_{k 1}^*,\dots,  \bphi_{k p}^*)^{\top}$ and $\bga_{k}^*=(\bga_{k 1}^*,\dots,  \bga_{k p}^*)^{\top}$ are the parameters for cluster $k$, $\delta_{(\bphi_{k}^{*}, \bga_{k}^{*})}(\cdot)$ is a Dirac measure at $(\bphi_{k}^{*}, \bga_{k}^{*})$, $\pi_{k}$ is the probability mass at atom $(\bphi_{k}^{*}, \bga_{k}^{*})$, and $K$ is a finite truncation for the maximum number of clusters. 
	The equation in~\eqref{eq:pyp1} implies that the model in~\eqref{eq:vcmm2} explores latent subpopulations by limiting the maximum number of subpopulations to $K$, not to infinity; see \cite{ishw:jame:01} for theoretical
	arguments.	
	  Meanwhile, the set of cluster-level parameters, $\bphi_{k}^*$ and $\bga_{k}^*$, is drawn from the base distribution $\mathcal H_{0}$, and the random weight $\pi_{k}$ derives from a set of random variables that each follows a beta distribution, i.e.,
	\begin{align*}	
		\pi_{k}~=~\pi_{k}(\V) ~=~V_{k}\prod_{\ell<k}(1-V_{\ell}),\quad V_{k}~\ind~{\rm Beta}\left(1, \nu \right),\quad k=1,\dots, K-1,
	\end{align*}
	where $\V=\{V_1,\dots,V_K\}$ and $V_{K}=1$, which guarantees the sum of all random weights is equal to one. Then, we can write $P(C_i=k|\V)=\pi_{k}(\V)$.
	The specification of $\nu$ may affect clustering performance. As $\nu$ goes to 0, the concentration toward the existing clusters gets stronger by decreasing the probability that a vector $(\bphi_{i}, \bga_{i})$ forms a new cluster. In this work, we set $\nu = 1$ by default so that every subject has the equal probability for shaping a new cluster.
	
	\subsection{Prior Specification for the Submodel Parameters}
	\label{sec:3.1}
	This section discusses the specification of prior distributions of each model component. The set of indicator variables $\bga_{kl}^{*}$ has a beta-binomial prior distribution, i.e.,
	\begin{align}
		p(\bga_{kl}^{*})~\propto~B(|\bga_{kl}^{*}|+a, M_{l}+1-|\bga_{kl}^{*}|+b), \label{gamma}
	\end{align}
	for $k=1,\dots, K$ and $l=1,\dots, p$, where $B(\cdot, \cdot)$ denotes the beta function. If  $a=b=1$, this prior distribution allocates equal probabilities for the number of active knots \citep{scot:etal:10}. This choice has been shown work successfully for function estimation with knot selection \citep{jeon:park:16,jeon:etal:17}.
	
	For the basis coefficients $\bphi_{(\bga_{k}^{*})}^{\star}$ that have varying dimension in each iteration,
	we consider the following mixture of $g$-priors,
	\begin{align}
		\begin{split}
		\bphi_{(\bga_{k}^{*})}^{\star}|(\bga_{k}^{*}, \tau_{k})~&\ind~{\rm N}_{|\bga_{k}^{*}|}\left(\mathbf{0}, \tau_{k}\R_{k(\bga_{k}^{*})}^{-1}\right),\\
\R_{k(\bga_{k}^{*})}~&=~		\sum_{i=1}^{N}\W_{i(\bga_{k}^{*})}^{\star\T}\W_{i(\bga_{k}^{*})}^{\star},\quad  k=1,\dots, K,  \\
\tau_{k}~&\iid{\rm IG}(1/2, N/2),\quad k = 1\dots, K.
\label{phi_star_prior}
		\end{split}
	\end{align}
		By characterizing the scale parameter with the total number of subjects, the prior in~(\ref{phi_star_prior}) corresponds to a Zellner-Siow prior, which is a multivariate Cauchy prior marginally for $\bphi_{(\bga_{k}^{*})}^{\star}$ \citep{lian:etal:08}. Therefore, the base measure $\mathcal H_0$ is constructed by combining a beta-binomial distribution and a multivariate Cauchy distribution.

The prior in \eqref{phi_star_prior} has several desirable properties. First, the prior distribution is invariant to linear transformations of the design matrix \citep{zell:86}. This means that the posterior distribution of the varying-coefficients is not affected by linear transformations of the basis functions in \eqref{eq:basis}. More importantly, the prior in \eqref{phi_star_prior} utilizes the population-level covariance, which is determined by assuming that all individuals belong to the $k$th cluster. This specific structure enhances the convergence efficiency of MCMC because the cluster-level basis coefficients $\bphi_{(\bga_{k}^{*})}^{\star}$ of empty clusters are sampled by taking advantage of the summed information of all subjects. As a result, the prior can naturally begin a new cluster to which a few subjects may belong.
 Indeed, it is well known that choosing a reasonable scale for the base measure $\mathcal H_0$ is very important in using the DP prior \citep{gelm:etal:15}. 
In this regard, our prior construction in \eqref{phi_star_prior} has a clear advantage over the related studies, which use the DP prior but do not fully account the concern of scale \citep{ray:mall:06,rodr:duns:14}.
Furthermore, having the right scale may be difficult with other penalty priors. For example, the Bayesian P-spline is a widely used Bayesian approach for nonparametric regression \citep{lang:brez:04}. Since its covariance structure plays an important role in smoothing, however, the prior distribution cannot be simply modified to have a reasonable scale for the DP prior.

	The remaining specification of priors for the fixed-dimensional parameter is standard.
	As for the fixed effects $\bbe$, we assign a multivariate normal distribution whose covariance matrix is positive definite, i.e.,
	\begin{align*}
		\bbe~\sim~{\rm N}_{q}(\mathbf{0},\PP).
	\end{align*}
For practical purposes, $\PP$ can be chosen as a diagonal matrix with large diagonal entries.
	For random effects, a multivariate normal distribution is used to generate the effects,
	\[
	\b_{i}|\bPsi ~\iid~ {\rm N}_{r}(\mathbf{0}, \bPsi),\quad i=1,\dots, N,
	\]
	where $\bPsi$ is the covariance matrix of the random effects and has an inverse-Wishart prior,
	\[
	\bPsi~\sim~ {\rm IW}(u, \D).
	\]
	In our study, $u$ and $\D$ are fixed in advance to make the prior distribution diffuse. 
	
	\subsection{Partially Collapsed Gibbs Sampler}
	\label{sec:pcg}

	Given the prior distributions in Section~\ref{sec:3.1}, we propose a sampling algorithm used to simulate the target posterior distribution,  
	\begin{align}
		p(\bphi_{(\bga^{*})}^{\star}, \bga^{*}, \bbe, \b, \C, \V, \btau, \bPsi, \Li|\Y),
		\label{eq:target}
	\end{align}
	where $\bga^{*}=\{\bga^{*}_{1},  \dots,\bga^{*}_{K}\}$, $\bphi_{(\bga^{*})}^{\star}=\{ \bphi_{(\bga_{1}^{*})}^{\star},  \dots,\bphi_{(\bga_{K}^{*})}^{\star}\}$, $\b=\{\b_{1}, \dots,\b_{N}\}$, $\btau=\{\tau_1, \dots,\tau_K\}$, $\C=\{C_1,\dots,C_N\}$, $\Li=\{\Li_{1},  \dots,\Li_{N}\}$, and $\Y=\{\Y_{1},  \dots,\Y_{N}\}$  denoting $\Y_{i}$ as a set of binary responses for subject $i$. To simulate the target posterior distribution in~(\ref{eq:target}), a standard Gibbs sampler based on its full conditional distributions cannot be implemented because the dimension $|\bga^{*}_{k}|\times 1$ of $\bphi_{(\bga^{*})}^{\star}$ depends on another model component $\bga^{*}$. In such a varying-dimensional problem, PCG sampling avoids the need of jumping between spaces of different dimensions through marginalization, permutation, and trimming, thereby making it implementable with the expectation of faster convergence; refer to Section~4 in \citet{park:vand:09}.
	In this study, we consider marginalizing the random effects $\b$ and the basis coefficients $\bphi_{(\bgam^*)}^{\star}$ in~\eqref{eq:target}, thereby producing the following marginal distributions,
	\begin{align}
		&p(\bga^{*}, \bbe, \C, \V, \btau, \bPsi, \Li|\Y), \label{eq:pcg1}\\
		&p(\bphi_{(\bga^{*})}^{\star}, \bga^{*}, \bbe, \C, \V, \btau, \bPsi, \Li| \Y). 
		\label{eq:pcg2}
	\end{align}

\begin{algorithm}[!t]
	\caption{One iteration of the PCG sampler}\label{pcg.algorithm}
	\KwData{$(\bphi_{(\bga^{*})}^{\star}, \bga^{*}, \bbe, \b, \C, \V, \btau, \bPsi, \Li)$}
	\For{$k=1,2,\dots,K$}{
		\For{$l=1,2,\dots,p $}{
			{\bf Step 1:} Draw $\bga_{kl}^{*}$ from $p(\bga_{kl}^{*}|\bga_{-kl}^{*},\bbe, \C, \V, \btau, \bPsi, \Li, \Y)$
		}
	}
	\For{$k=1,2,\dots,K$}{
		{\bf Step 2:} Draw $V_{k}$ from $p(V_k|\bga^{*}, \bbe, \C, \btau, \bPsi, \Li, \Y)$
	}
	\For{$k=1,2,\dots,K$}{
		{\bf Step 3:} Draw $\bphi_{(\bga_{k}^{*})}^{\star}$ from $p(\bphi_{(\bga_{k}^{*})}^{\star}|\bga^{*}, \bbe, \C, \V, \btau, \bPsi, \Li, \Y)$
	}
	{\bf Step 4:} Draw $\bbe$ from $p(\bbe|\bphi_{(\bga^{*})}^{\star}, \bga^{*}, \C, \V,  \btau, \bPsi, \Li, \Y)$,
	
	\For{$i=1,2,\dots,N$}{
		{\bf Step 5:} Draw $\b_{i}$ from $p(\b_i|\bphi_{(\bga^{*})}^{\star}, \bga^{*}, \bbe, \C, \V, \btau, \bPsi, \Li, \Y)$
	}
	\For{$k=1,2,\dots,K$}{
		{\bf Step 6:} Draw $\tau_{k}$ from $p(\tau_k|\bphi_{(\bga^{*})}^{\star}, \bga^{*}, \bbe, \b, \C, \V, \bPsi, \Li, \Y)$
	}				
	{\bf Step 7:} Draw $\bPsi$ from  $p(\bPsi|\bphi_{(\bga^{*})}^{\star}, \bga^{*}, \bbe, \b, \C, \V, \btau, \Li, \Y)$,
	
	\For{$i=1,2,\dots,N$}{
		\For{$j=1,2,\dots,n_{i}$}{
			{\bf Step 8:} Draw $L_{ij}$ from $p(L_{ij}|\bphi_{(\bga^{*})}^{\star}, \bga^{*}, \bbe, \b, \C, \V, \btau, \bPsi, \Y)$
		}
	}
	\For{$i=1,2,\dots,N$}{
		{\bf Step 9:} Draw $C_{i}$ from $p(C_i|\bphi_{(\bga^{*})}^{\star}, \bga^{*}, \bbe, \b, \V, \btau, \bPsi, \Li, \Y)$
	}
\end{algorithm}

	One iteration of the PCG sampler is shown in Algorithm~\ref{pcg.algorithm}. Steps 1 and 2 are marginalized by using \eqref{eq:pcg1}, while Steps 3 and 4 are marginalized by using \eqref{eq:pcg2}. To maintain the target stationary distribution of
	Algorithm~\ref{pcg.algorithm}, the sampling steps are permuted in a specific order. Trimming is used to remove the redundant samples of components. For more applications of the PCG sampling including other varying dimensional cases, refer to \citet{park:vand:09,jeon:park:16,jeon:etal:17,park:jeon:17,park:etal:19}. Because the target stationary distribution of Algorithm~\ref{pcg.algorithm} is maintained in a specific order, the change of the order of sampling steps may not guarantee the stationarity of a Markov chain, and care must be taken not to change the sampling order; refer to \cite{vand:park:08}. The details of Algorithm~\ref{pcg.algorithm} are given in \ref{sec:appen}.

	\section{A Simulation Study}
	\label{sec:sim}

In this section, we validate the robustness and sensitivity of the proposed method through extensive simulation studies. All simulation results are based on 300 replicated datasets. 

	\subsection{Simulation Setting}
	\label{sec:setting}
	
	Throughout the simulation, we mainly consider three groups of varying coefficients, and the total number $N$ of subjects and the number of subjects in each cluster will be specified later based on simulation setups. The values of an underlying effect modifier~$t$, $\{t_{ij}:i=1,\dots,N, j=1,\dots,n_{i}\}$, are randomly generated from a uniform distribution between 0 and 1, and the values of known covariates for subject~$i$, i.e., $\W_{i}$, $\X_{i}$, and $\Z_{i}$, are independently generated from a standard normal distribution, except that the first column of both $\W_{i}$ and $\Z_{i}$ is set to a column vector of 1's. Within the range of $t$ between 0 and 1, $\alpha_{kl}(t)$ denotes a varying-coefficient function of the $l$th covariate in cluster $k$. The varying coefficients for three clusters are constant, linear, or nonlinear, as described below: 
	\begin{align*}
		\alpha_{11}(t) ~& =~  2\exp\{-200(t-0.2)^{2}\}+\exp\{-10(t-0.6)^{2}\},\\
		\alpha_{12}(t) ~& =~  \sin(2\pi t^3),\\
		\alpha_{21}(t) ~& =~  \sin\{8(t-0.5)\}+1.5\exp\{-400(t-0.5)^2\},\\
		\alpha_{22}(t) ~& =~  2t, \\
		\alpha_{31}(t) ~& =~ -2t,\\
		\alpha_{32}(t) ~& =~ 0.
	\end{align*}
The true values of the other model parameters are set to $\bbe=(\beta_1,\beta_2)^{\top}=(1,-1)^{\top}$ and 
	\begin{align*}
		\bPsi~=~\left(\begin{array}{cc}
			\psi_{11} & \psi_{12}\\
			\psi_{12} & \psi_{22}
		\end{array}\right) ~=~ \left(\begin{array}{cc}
				0.5 & 0.25\\
				0.25 & 0.8
		\end{array}\right).
	\end{align*}
The latent response $L_{ij}$ of the probit varying-coefficient mixed model is drawn from
	\begin{align*}
		&L_{ij} ~\sim~ {\rm N}\left(\W_{i}^{ (j)}\bal_{C_{i}}(t_{ij}) + \X_{i}^{(j)}\bbe +  \Z_{i}^{(j)}\b_{i}, 1\right), \quad i=1,\dots, N,\quad j=1,\dots, n_{i},
	\end{align*}
	where $\bal_{C_{i}}(t_{ij})=(\alpha_{C_i 1}(t_{ij}),\alpha_{C_i 2}(t_{ij}))^{\top}$ and it is used to generate a series of binary data such that $Y_{ij} = I(L_{ij}>0)$ for observation $j$ on subject $i$.

	\subsection{Performance of the Proposed Method}
	\label{sec:replicates}

In this section, we demonstrate the performance of the proposed method under various simulation setups. We consider three different scenarios: Scenario~I, where each cluster has 400 subjects ($N = 1200$), Scenario~II, where each cluster has 200 subjects ($N = 600$), and Scenario~III, where the three clusters have 600, 400, and 200 subjects, respectively ($N = 1200$). For each scenario, three different values of the concentration parameter are also considered to examine the robustness of the DP prior: $\nu\in\{0.1,1,10\}$. The proposed method is applied to each combination of the scenarios for the sample size and the concentration parameter with 300 replications of the datasets. We run 20,000 iterations of the PCG sampler, discarding the first half of the draws as burn-in and using the second half for our posterior analysis.

\begin{figure}[!t]
	\centering
	\psfrag{p}[][][0.8]{Precision}
	\psfrag{r}[][][0.8]{Recall}		
	\psfrag{f}[][][0.8]{F1-score}		
	
	\includegraphics[width=0.60\textheight]
	{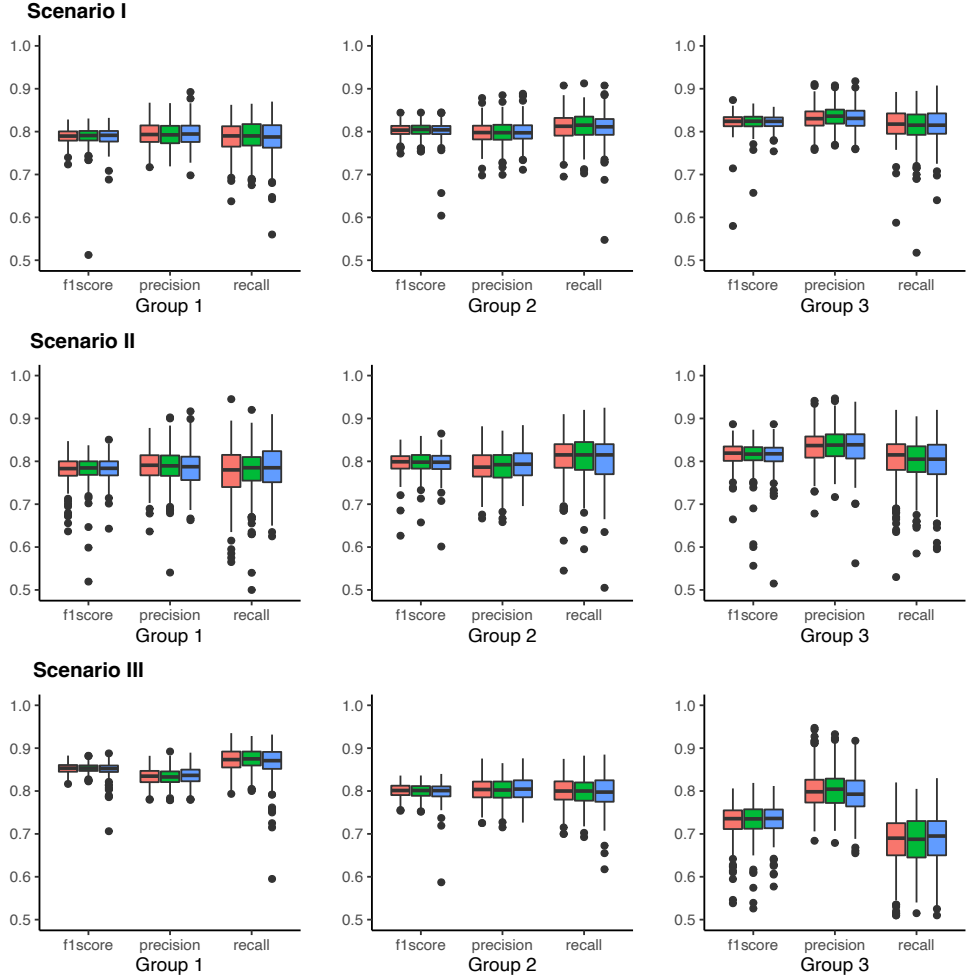}
	\caption
	{Performance measures for the clustering procedure obtained from 300 replicated datasets with $\nu=0.1$ (red), $\nu=1$ (green), and $\nu=10$ (blue).}
	\label{simul:cluster_metrics}
\end{figure}

\begin{figure}[p]
	\centering
	\psfrag{a11}[][][0.6]{$\alpha_{11}(\cdot)$}
	\psfrag{a12}[][][0.6]{$\alpha_{12}(\cdot)$}
	
	\psfrag{a21}[][][0.6]{$\alpha_{21}(\cdot)$}
	\psfrag{a22}[][][0.6]{$\alpha_{22}(\cdot)$}
	
	\psfrag{a31}[][][0.6]{$\alpha_{31}(\cdot)$}
	\psfrag{a32}[][][0.6]{$\alpha_{32}(\cdot)$}
	
	\includegraphics[width=0.60\textheight]
	{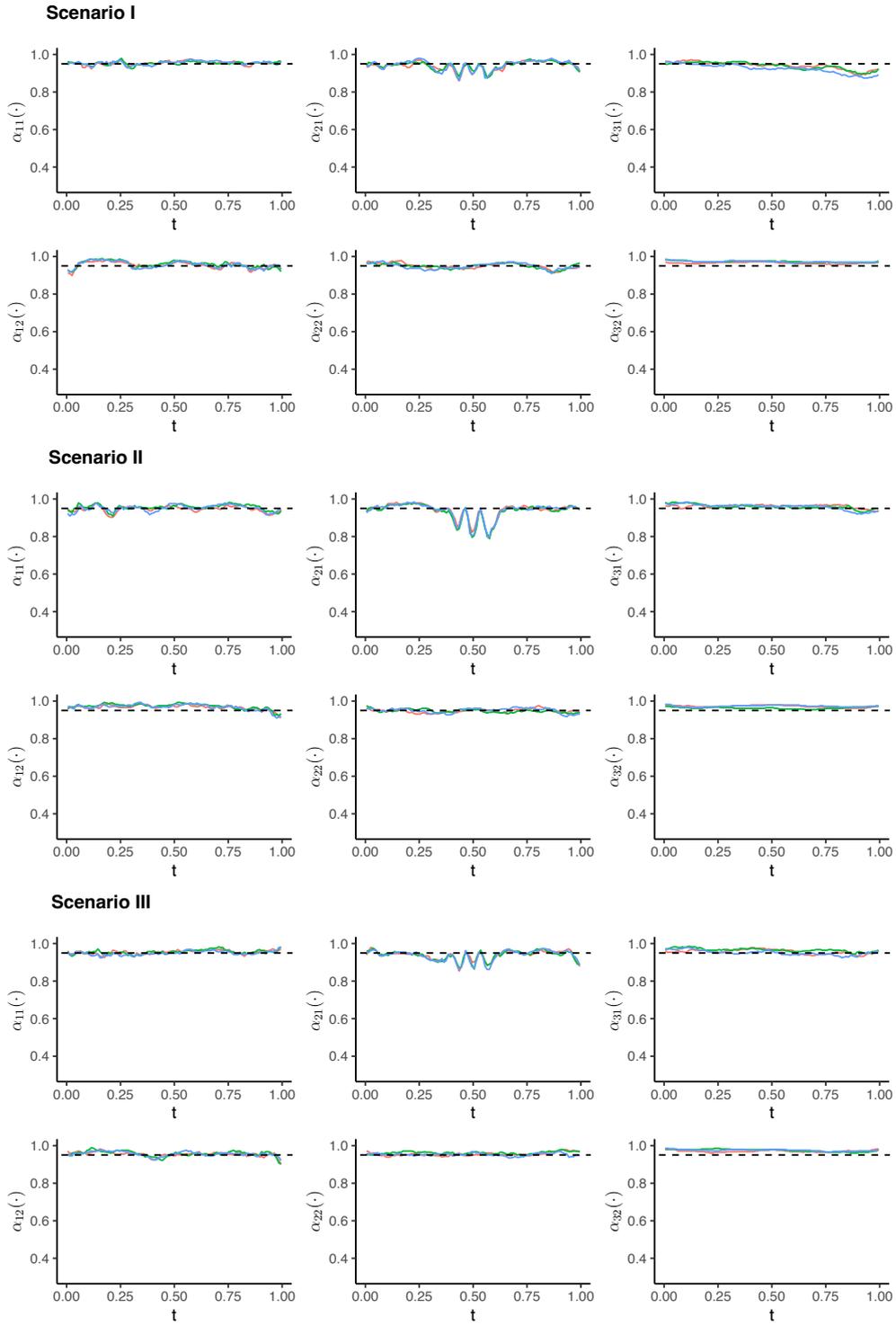}
	\caption
	{Coverage probabilities of the pointwise 95\% credible bands for the varying coefficients obtained from 300 replicated datasets with $\nu=0.1$ (red), $\nu=1$ (green), and $\nu=10$ (blue).}
	\label{simul:covprob_var}
\end{figure}

\begin{table}[!t]
	{ \caption{RMSE of the posterior median and coverage probabilities of the 95\% credible intervals for the fixed-dimensional parameters obtained from 300 replicated datasets.}
		\begin{centering}
			\label{simul:covprob_fix}  %
			\begin{tabular}{rrrrrrrr}
				\toprule
				& & \multicolumn{3}{c}{RMSE} & \multicolumn{3}{c}{95\% coverage}  \\
				\cmidrule(lr){3-5} \cmidrule(lr){6-8} 
				&   Parameter & $\nu=0.1$  & $\nu=1$  & $\nu=10$ & $\nu=0.1$  & $\nu=1$  & $\nu=10$  \\
				\midrule
				Scenario I & $\beta_1$    & 0.029 & 0.030 & 0.031 & 0.930 & 0.930 & 0.930   \\
				           & $\beta_2$  & 0.032 & 0.033 & 0.033 & 0.913 & 0.920 & 0.909   \\
            			   & $\psi_{11}$   & 0.070 & 0.072 & 0.072 & 0.937 & 0.916 & 0.937   \\
				           & $\psi_{12}$  & 0.042 & 0.043 & 0.043 & 0.955 & 0.944 & 0.948   \\
				           & $\psi_{22}$ & 0.073 & 0.074 & 0.075 & 0.937 & 0.934 & 0.941   \\
				\midrule
				Scenario II & $\beta_1$    & 0.042 & 0.046 & 0.044 & 0.962 & 0.919 & 0.920   \\
				            & $\beta_2$    & 0.041 & 0.047 & 0.046 & 0.941 & 0.912 & 0.934   \\
				            & $\psi_{11}$  & 0.119 & 0.124 & 0.121 & 0.882 & 0.908 & 0.885   \\
				            & $\psi_{12}$  & 0.067 & 0.069 & 0.069 & 0.948 & 0.951 & 0.944   \\
				            & $\psi_{22}$  & 0.102 & 0.110 & 0.107 & 0.948 & 0.930 & 0.948   \\
				\midrule
				Scenario III & $\beta_1$     & 0.029 & 0.030 & 0.031 & 0.946 & 0.932 & 0.917   \\
				             & $\beta_2$     & 0.032 & 0.031 & 0.034 & 0.907 & 0.929 & 0.897   \\
				             & $\psi_{11}$   & 0.063 & 0.064 & 0.064 & 0.929 & 0.943 & 0.921   \\
				             & $\psi_{12}$   & 0.041 & 0.041 & 0.042 & 0.946 & 0.957 & 0.952   \\
				             & $\psi_{22}$   & 0.077 & 0.077 & 0.079 & 0.946 & 0.954 & 0.941   \\
				\bottomrule
			\end{tabular}
			\par\end{centering}
	}
\end{table}	

The side-by-side boxplots in Figure~\ref{simul:cluster_metrics} illustrate clustering performance for the 300 replicated datasets in terms of precision, recall, and F1-score. The clustering labels of all subjects are chosen by the posterior modes. As expected, clusters with larger sample sizes perform better in clustering. The metrics show similar results across different values of the concentration parameter $\nu$, indicating that clustering performance is robust to the specification of the hyperparameter for the DP prior. We calculate coverage probabilities of the pointwise 95\% credible bands with the 300 replicates, where the bands are specified by the 2.5\% and 97.5\% posterior quantiles. Figure~\ref{simul:covprob_var} shows that the coverage probabilities are close to the nominal value of $0.95$, which validates the uncertainty quantification through the posterior distribution. The coverage probabilities are consistent for different values of $\nu$, further supporting the robustness of our proposed method against the hyperparameter specification. With 300 replicated datasets, Table~\ref{simul:covprob_fix} shows the root-mean-square errors (RMSE) of the posterior median and coverage probabilities of the 95\% credible intervals for the fixed-dimensional parameters. The results indicate that the fixed-dimensional parameters are also insensitive to the hyperparameter specification.

\begin{figure}[!t]
	\centering
	\psfrag{nu}[][][0.6]{$\nu$}
	
	\includegraphics[width=0.60\textheight]
	{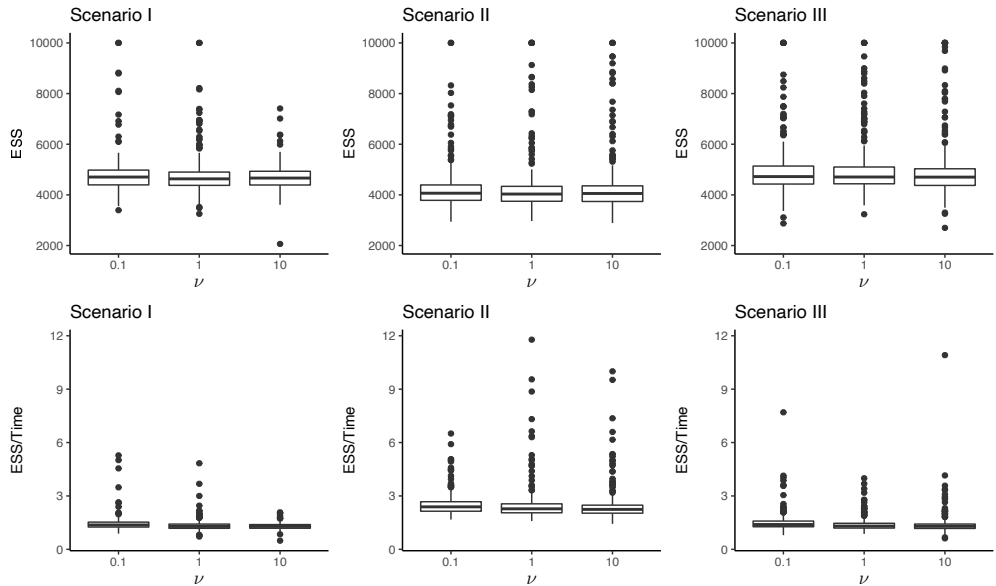}
	
	\caption
	{Estimates of the multivariate ESS out of 10,000 PCG iterations and the multivariate ESS per second. The estimates of the multivariate ESS are replaced by 10,000 if they are larger than 10,000. }
	\label{simul:ess}
\end{figure}

To assess the efficiency of our proposed PCG sampler, we calculate the multivariate effective sample size (ESS) \citep{vats:etal:19} for all simulation settings, as shown in Figure~\ref{simul:ess}. The second half of the chain with 20,000 iterations is used to calculate the multivariate ESS, along with the running time of the sampler on a server equipped with CentOS7 and two Sky Lake CPUs @ 2.60GHz. 
The target parameters have a dimension of 185, including two varying-coefficient functions with 30 knots for each of three clusters and five fixed-dimensional parameters. As shown in Figure~\ref{simul:ess}, the multivariate ESS is approximately 4,000--5,000 out of 10,000 iterations, implying that the proposed sampling algorithm exhibits reasonable convergence characteristics. Figure~\ref{simul:ess} also shows the multivariate ESS divided by time (seconds), providing the number of independent draws obtained per unit time. The results demonstrate that roughly 1.5 to 2.5 independent draws are obtained per second.
Furthermore, the concentration hyperparameter has no significant influence on the multivariate ESS, indicating the robustness of our proposed method to the specification of the hyperparameter. 


\begin{figure}[p]
	\centering
	\psfrag{v11}[][][0.6]{$\alpha_{11}(\cdot)$}
	\psfrag{v12}[][][0.6]{$\alpha_{12}(\cdot)$}
	
	\psfrag{v21}[][][0.6]{$\alpha_{21}(\cdot)$}
	\psfrag{v22}[][][0.6]{$\alpha_{22}(\cdot)$}

	\psfrag{v31}[][][0.6]{$\alpha_{31}(\cdot)$}
	\psfrag{v32}[][][0.6]{$\alpha_{32}(\cdot)$}
	
	\psfrag{t}[][][0.6]{$\textit{t}$}
	
	\includegraphics[width=0.6\textheight]{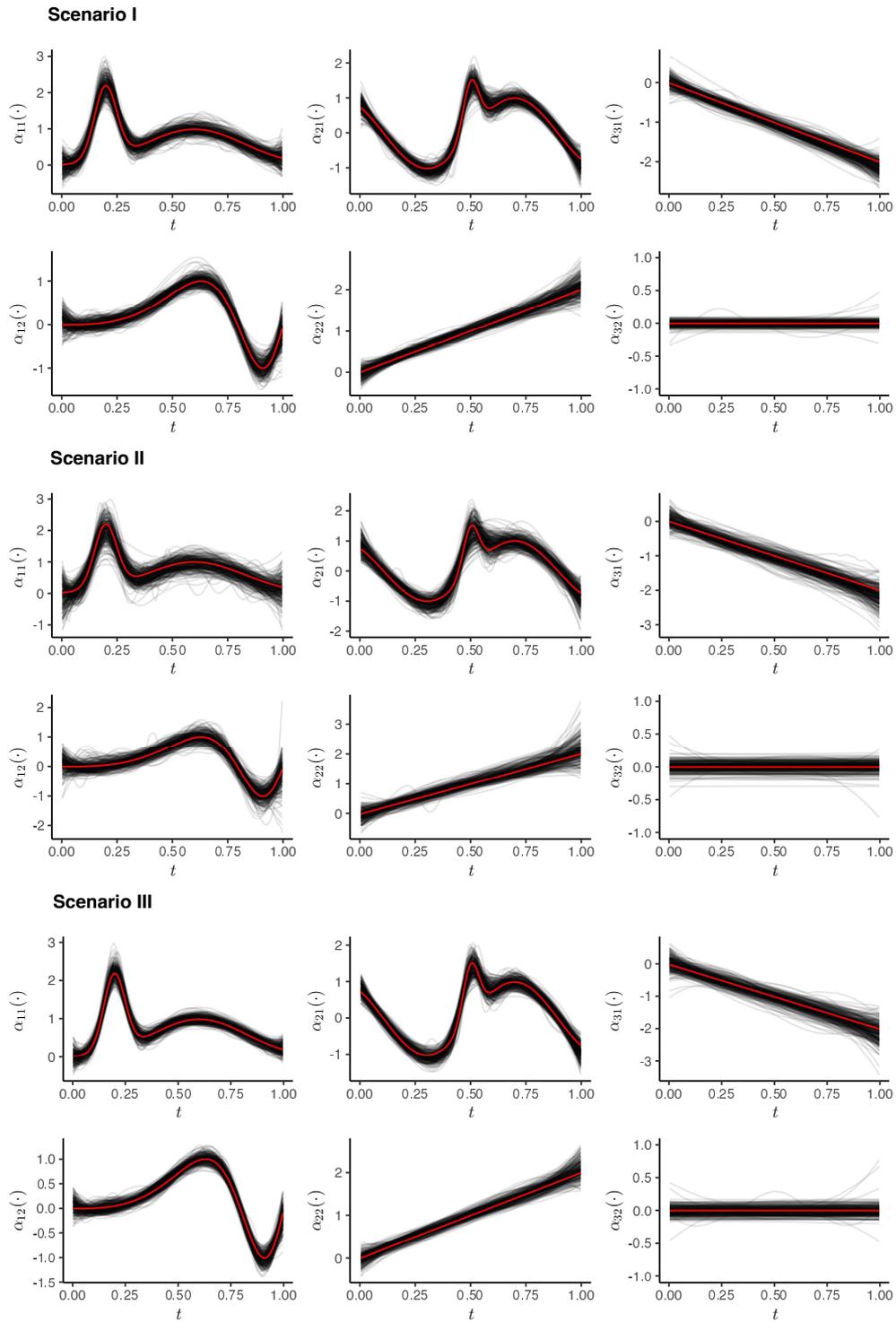}
	\caption
	{Pointwise posterior medians of the varying coefficients for 300 replicated datasets (gray solid lines) and the true varying-coefficient functions (red solid lines).}
	\label{simul:combined_vc_plots}
\end{figure}

Figure~\ref{simul:combined_vc_plots} overlays the pointwise posterior medians of the varying coefficients of the 300 replicated datasets using the default concentration parameter value of $\nu=1$.  Our results show that the estimated posterior medians become closer to the true functions as sample sizes increase. We also examined the posterior medians obtained with alternative values of~$\nu$, specifically $\nu=0.1$ and $\nu=10$, but found that the results were similar and therefore omitted here.

	\subsection{Effects of Ignoring Subpopulation}
	\label{sec:comp_jeong}

To evaluate the necessity of subpopulation modeling, we compared our proposed method with two competing approaches that do not account for functional clustering: \citet{jeon:etal:17} and the \texttt{mgcv} package \citep{wood:17}. \citet{jeon:etal:17} is a fully Bayesian method for estimating probit varying-coefficient mixed models with a homogeneous population using free-knot splines \citep{smit:kohn:96}. The \texttt{mgcv} package uses penalized quasi-likelihood to estimate the same model. To avoid redundancy, we compared the two competing methods with our proposed method using replicated datasets under Scenario~I specified in Section~\ref{sec:replicates}

\begin{figure}[!t]
	\centering
	
	\psfrag{j1}[][][0.5]{$\alpha_{\cdot 1}(\cdot)$}
	\psfrag{j2}[][][0.5]{$\alpha_{\cdot 2}(\cdot)$}
	
	\psfrag{m1}[][][0.5]{$\alpha_{\cdot 1}(\cdot)$}
	\psfrag{m2}[][][0.5]{$\alpha_{\cdot 2}(\cdot)$}
	
	\psfrag{a}[][][0.6]{$\textit{t}$}
	
	\includegraphics[width=0.50\textheight]{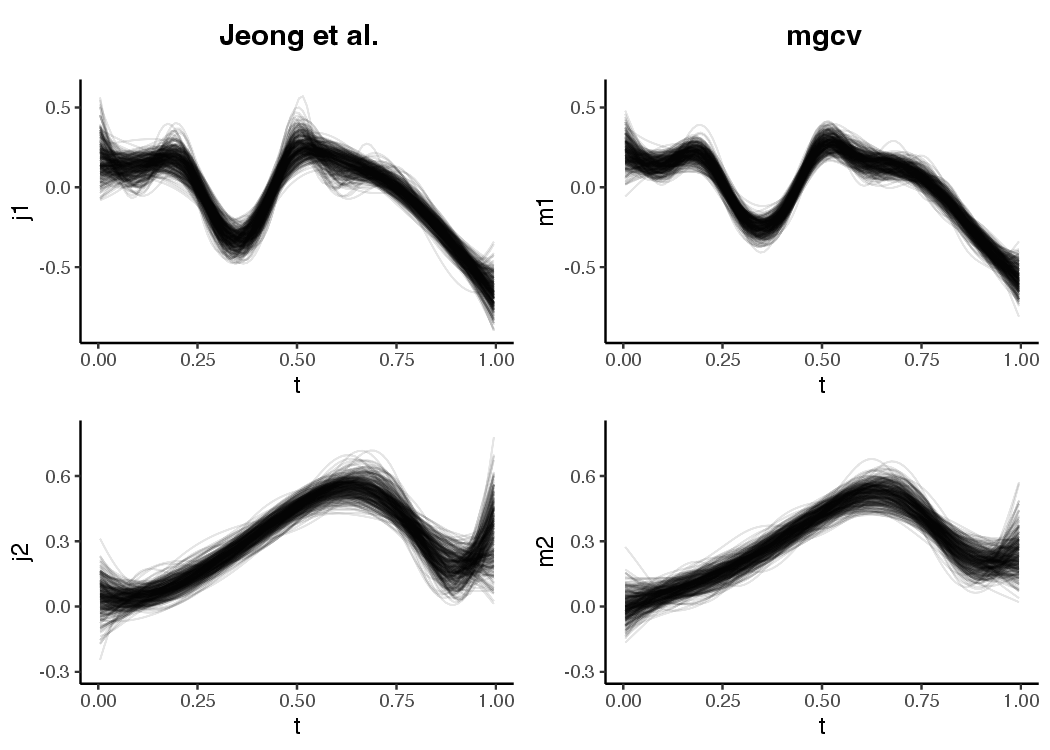}
	
	\caption{Estimates of the varying coefficients obtained from 300 replicated datasets. The estimates are the pointwise posterior medians for \citet{jeon:etal:17} and the penalized quasi-likelihood estimates for \texttt{mgcv}.
	}
	\label{simul:comp_var_hete}
\end{figure}

\begin{table}[!t]
	\caption{RMSE of the estimates and coverage probabilities of the 95\% intervals for the fixed-dimensional parameters obtained from 300 replicated datasets. The coverage probabilities are obtained by the 95\% credible intervals for \citet{jeon:etal:17} and the 95\% confidence intervals approximated with standard errors for \texttt{mgcv}. The \texttt{mgcv} package does not provide standard errors for the covariance of random effects.}
	\label{simul:comp_fix}
	\centering
	\begin{tabular}{rrrrrrrrr}
		\toprule
		& \multicolumn{2}{c}{Proposed method ($\nu=1$)}  & \multicolumn{2}{c}{\citet{jeon:etal:17}} &  \multicolumn{2}{c}{\texttt{mgcv}}  \\
		\cmidrule(lr){2-3} \cmidrule(lr){4-5} \cmidrule(lr){6-7} 
		Parameter & RMSE & 95\% coverage & RMSE & 95\% coverage & RMSE & 95\% coverage \\
		\midrule
		$\beta_1$ & 0.030 & 0.930 &  0.205 & 0.000 &  0.190 &   0.000  \\
		$\beta_2$ & 0.033 & 0.920 & 0.204 & 0.000 &  0.190 &  0.000  \\
		$\psi_{11}$ & 0.072 & 0.916 & 0.220 & 0.000 & 0.199 & -  \\
		$\psi_{12}$ & 0.043 & 0.944 & 0.097 & 0.170 & 0.099 & - \\
		$\psi_{22}$ & 0.074 & 0.934 & 0.299 & 0.000 & 0.351 & -  \\
		\bottomrule
	\end{tabular}
\end{table}

In Figure~\ref{simul:comp_var_hete}, we present the estimates of varying coefficients obtained by two competing methods that do not account for functional clustering. The estimated trends appear to be the average of the varying coefficients across the clusters in Scenario~I, resulting in significant estimation bias for the cluster-specific effects. Furthermore, ignoring subpopulations leads to significant estimation bias for the fixed-dimensional parameters, as demonstrated in Table~\ref{simul:comp_fix}, despite the fact that they are common across all clusters. This suggests that accounting for subpopulation effects is crucial even for parameters that are shared among clusters.


\subsection{Homogeneous Population}	
\label{sec:homopop}

We also investigate the performance of the proposed method in a homogeneous population with a single true cluster. To generate the simulation datasets, we set all subjects to have the same values of $\alpha_{11}(t)$ and $\alpha_{12}(t)$, as described in Section~\ref{sec:setting}. Specifically, we set $\bal_{C_i}(t_{ij})=(\alpha_{11}(t_{ij}),\alpha_{12}(t_{ij}))^{\top}$ for all $i$, resulting in a single cluster. The number of subjects is set to $N=300$, and all other simulation settings are identical to those specified in Section~\ref{sec:setting}.

\begin{figure}[!t]
	\centering

	\psfrag{v11}[][][0.5]{$\alpha_{11}(\cdot)$}
	\psfrag{v12}[][][0.5]{$\alpha_{12}(\cdot)$}

	\psfrag{j1}[][][0.5]{$\alpha_{11}(\cdot)$}
	\psfrag{j2}[][][0.5]{$\alpha_{12}(\cdot)$}
	
	\psfrag{m1}[][][0.5]{$\alpha_{11}(\cdot)$}
	\psfrag{m2}[][][0.5]{$\alpha_{12}(\cdot)$}
	
	\psfrag{t}[][][0.6]{$\textit{t}$}
	
	\includegraphics[width=0.60\textheight]{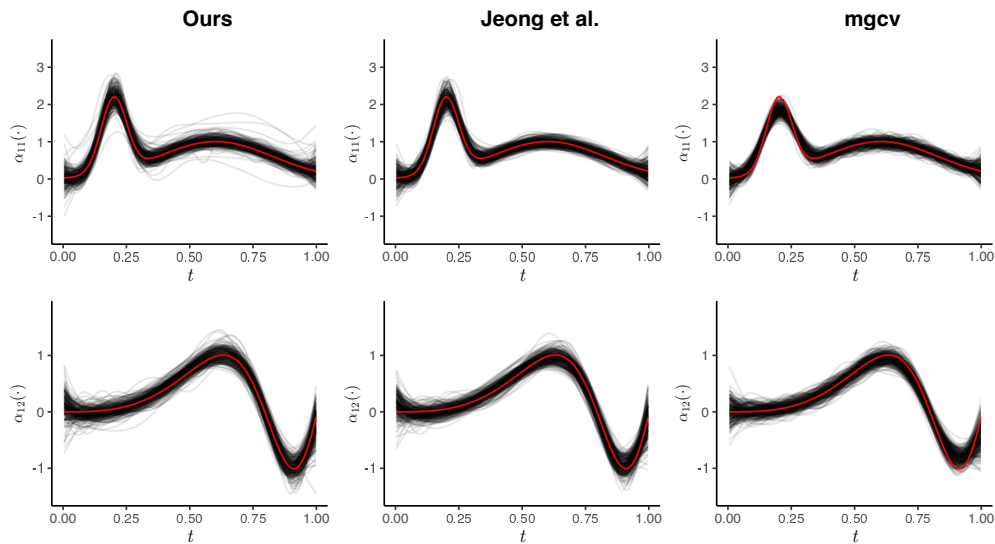}
	
	\caption{Estimates of varying coefficients of 300 replications (solid gray lines) and the true functions (solid red lines). The estimates are the pointwise posterior medians for the proposed method and \citet{jeon:etal:17} and the penalized quasi-likelihood estimates for \texttt{mgcv}.}
	\label{simul:comp_var_homo}
\end{figure}

\begin{figure}[!t]
	\centering
	
	\psfrag{a11}[][][0.5]{$\alpha_{11}(\cdot)$}
	\psfrag{a12}[][][0.5]{$\alpha_{12}(\cdot)$}
	
	\psfrag{t}[][][0.6]{$\textit{t}$}
	
	\includegraphics[width=0.55\textheight]{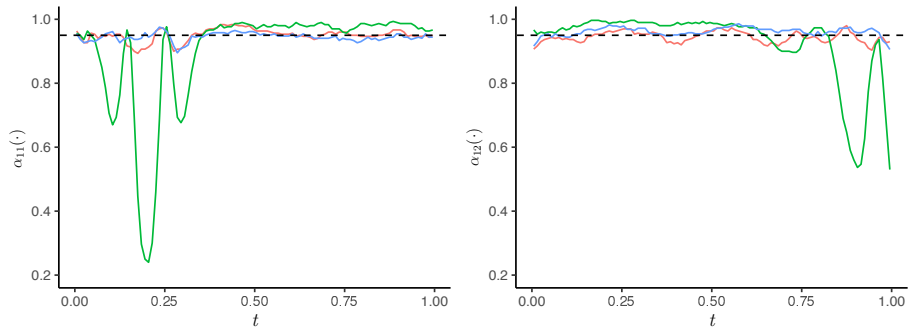}
	
	\caption{Coverage probabilities of the 95\% intervals for the varying coefficients obtained from 300 replicated datasets: the proposed method (red), \citet{jeon:etal:17} (blue), and \texttt{mgcv} (green). The coverage probabilities are obtained by the pointwise 95\% credible bands for the proposed method and \citet{jeon:etal:17} and the pointwise 95\% confidence bands approximated with standard errors for \texttt{mgcv}.}
	\label{simul:comp_var_homo_cov}
\end{figure}

\begin{table}[!t]
	\caption{RMSE of the estimates and coverage probabilities of the 95\% intervals for the fixed-dimensional parameters obtained from 300 replicated datasets. The coverage probabilities are obtained by the 95\% credible intervals for \citet{jeon:etal:17} and the 95\% confidence intervals approximated with standard errors for \texttt{mgcv}. The \texttt{mgcv} package does not provide standard errors for the covariance of random effects.}
	\label{simul:comp_fix_homo}
	\centering
	\begin{tabular}{rrrrrrrrr}
		\toprule
		& \multicolumn{2}{c}{Proposed method ($\nu=1$)}  & \multicolumn{2}{c}{\citet{jeon:etal:17}} &  \multicolumn{2}{c}{\texttt{mgcv}}  \\
		\cmidrule(lr){2-3} \cmidrule(lr){4-5} \cmidrule(lr){6-7} 
		Parameter & RMSE & 95\% coverage & RMSE & 95\% coverage & RMSE & 95\% coverage \\
		\midrule
		$\beta_1$ &  0.055  & 0.954 & 0.048 & 0.960 & 0.058 &  0.916  \\
		$\beta_2$ &  0.055  & 0.944 & 0.050 & 0.953 & 0.058 &  0.923  \\
		$\psi_{11}$ & 0.111 & 0.923 & 0.089 & 0.950 & 0.678 & -  \\
		$\psi_{12}$ & 0.073 & 0.958 & 0.066 & 0.960 & 0.093 & - \\
		$\psi_{22}$ & 0.143 & 0.944 & 0.135 & 0.943 & 0.148 & -  \\
		\bottomrule
	\end{tabular}
\end{table}

 Figure~\ref{simul:comp_var_homo} displays the pointwise posterior medians of the varying coefficients using the proposed method and two competitors under the homogeneous population assumption. The proposed method shows a reasonably small estimation bias compared to~\citet{jeon:etal:17} despite some deviation for the incorrectly clustered subjects, implying its applicability without knowing the population structure. In contrast, \texttt{mgcv} yields larger estimation bias than the other two methods. Figure~\ref{simul:comp_var_homo_cov} presents the coverage probabilities of the pointwise 95\% credible or confidence bands of the varying coefficients, further demonstrating the worse performance of \texttt{mgcv}. Table~\ref{simul:comp_fix_homo} summarizes the results of the fixed-dimensional parameters, indicating that the RMSEs of the proposed method are slightly larger than those for \citet{jeon:etal:17}. Considering the flexibility of the proposed method in accounting for potentially heterogeneous populations, however, it can be deemed more useful than \citet{jeon:etal:17}.


	\section{Application to Binary Longitudinal Data}
	\label{sec:app}
	\subsection{Data Description and Modeling Procedure}\label{sec:app_data}
	In this section, we consider the German Socioeconomic Panel (GSOEP) data \citep{riph:etal:03}. The dataset consists of repeated observations from 7,293 subjects in Germany for the years 1984--1988, 1991, and 1994. The response variable of interest is working status (employed=1; otherwise=0) and covariates consist of $A_{ij}$ (age), $M_{ij}$ (marital status; married=1, otherwise=0), $K_{ij}$ (children under the age of 16 in the household; yes=1, no=0), $H_{ij}$ (degree of handicap; 0 to 100 in percent), and $S_{ij}$ (personal health satisfaction; 0 to 10). We confine our samples to 893 subjects under the age of 53 with Abitur degrees in order to examine the varying effect of having young children in the household on working status as a function of age for people with secondary education.
	
According to our preliminary analysis, which assumes varying effects for all covariates, we have decided to treat handicap, personal health satisfaction, and marital status as constant effects in subsequent analyses. As a result, we aim to model the varying effects of the intercept and the presence of children under the age of 16, while treating the remaining covariates as fixed effects. Our target model is then given by 
	\begin{align*}
		L_{ij}~=~\alpha_{{C_{i}}0}^{*}(A_{ij})+\alpha_{{C_{i}}K}^{*}(A_{ij})K_{ij}+\beta_M M_{ij}+ \beta_{S} S_{ij} + \beta_{H} H_{ij} +b_{i}+\epsilon_{ij}. ~~~\label{eq:5.1}
	\end{align*}
The reduced model complexity based on the preliminary analysis leads to faster computation and greater stability.

	%
	%
	%
	
	Similar to Section~\ref{sec:comp_jeong}, we compare the proposed method with the model in \cite{jeon:etal:17} that ignores heterogeneity among subjects; \texttt{mgcv} is not considered because \cite{jeon:etal:17} outperforms it (see Section~\ref{sec:homopop}).
	 The corresponding simple model is given by 
	\begin{align*}
		L_{ij}~=~\alpha_{0}(A_{ij})+\alpha_{K}(A_{ij})K_{ij}+ \beta_M M_{ij} + \beta_{S} S_{ij} + \beta_{H} H_{ij} +b_{i}+\epsilon_{ij}.
	\end{align*}
	As shown in the next section, the simple model fails to account for heterogeneity among samples and can be obtained by pooling the results of the proposed target model.

	\subsection{Analysis and Results} \label{anal_res}

	\begin{figure}[!t]
		\centering
		\psfrag{b1}[][][0.6]{$\beta_{S}$}
		\psfrag{b2}[][][0.6]{$\beta_{H}$}
		\psfrag{b3}[][][0.6]{$\beta_{M}$}
		\psfrag{psi}[][][0.6]{$\psi$}
		
		\psfrag{a11}[][][0.6]{$\alpha^*_{G_1 0}(\cdot)$}
		\psfrag{a12}[][][0.6]{$\alpha^*_{G_1 K}(\cdot)$}	
		\psfrag{a21}[][][0.6]{$\alpha^*_{G_2 0}(\cdot)$}
		\psfrag{a22}[][][0.6]{$\alpha^*_{G_2 K}(\cdot)$}		
		
		\includegraphics[width=0.55\textheight]{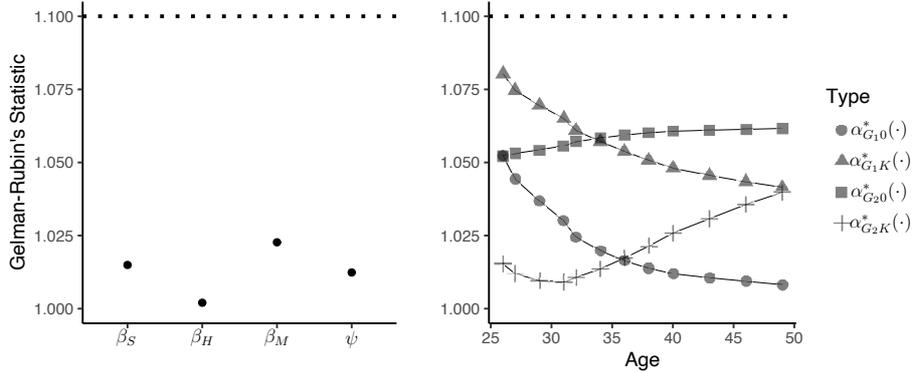}
		\caption
		{The $R^{1/2}$ statistics for the fixed-dimensional parameters (left) and the fixed points of the varying coefficients (right).}
		\label{real:convergence}
	\end{figure}
	
%
%
%
%
%
%
%
	
	\begin{table}[!t]
		{ \caption{Characteristics of the clustered groups}
			
			\label{Characteristics} \centering %
			\begin{tabular}{ccccc}
				\toprule
				~~~~~Covariate~~~~~~ & ~~\text{Overall}~~ & ~~\text{Group 1}~~  & ~~\text{Group 2}~~    \tabularnewline
				\midrule
				
				\text{Female} & 35.1\% & 33.3\%  & 89.1\%  
				\tabularnewline
				
				\text{Married} & 57.7\% & 55.9\%  & 71.2\% 
				\tabularnewline
				
				\text{White-collar workers} & 38.7\% & 41.1 \%  & 20.8\% 
				\tabularnewline
				
				\text{Civil servants} & 25.8\% & 29.0\%  & 4.0\%
				\tabularnewline
				
				\text{University} & 57.2\% & 58.4\% & 42.0\% 
				\tabularnewline
				
				\bottomrule
			\end{tabular}
			\par
		}
	\end{table}
	
	\begin{figure}[!t]
		\centering
		\psfrag{a11}[][][0.6]{$\alpha_{G_{1}0}^*(\cdot)$}				
		\psfrag{a21}[][][0.6]{$\alpha_{G_{2}0}^*(\cdot)$}
		\psfrag{a31}[][][0.6]{$\alpha_{G_{3}0}^*(\cdot)$}
		
		\psfrag{a12}[][][0.6]{$\alpha_{G_{1}K}^*(\cdot)$}				
		\psfrag{a22}[][][0.6]{$\alpha_{G_{2}K}^*(\cdot)$}
		\psfrag{a32}[][][0.6]{$\alpha_{G_{3}K}^*(\cdot)$}
		
		\psfrag{a}[][][0.6]{Age}

		\includegraphics[width=0.55\textheight]{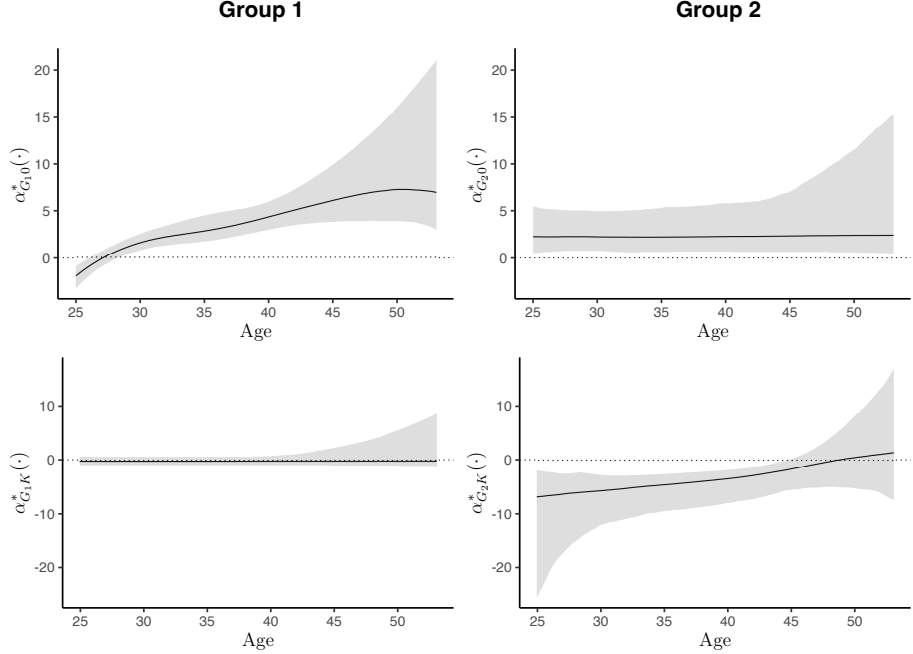}

		\caption
		{Posterior summaries of the varying coefficients for two major groups. Solid lines represent the pointwise posterior medians of each varying coefficient function and gray regions correspond to pointwise 95\% posterior intervals.}
		\label{univ:real_varying}
	\end{figure}

\begin{table}[!t]
	{ \caption{Posterior summaries of the fixed-dimensional parameters}
		\begin{centering}
			\label{summary}  %
			\begin{tabular}{lrrrrrr}
				\toprule
				& \multicolumn{3}{c}{Proposed method} & \multicolumn{3}{c}{\citet{jeon:etal:17}} \\
				\cmidrule(lr){2-4} \cmidrule(lr){5-7}
				Parameter  & Mean  & Median  & 95\% interval  & Mean  & Median  & 95\% interval \\
				\midrule
				$\psi$  & $3.539$  & $3.357$  & $(1.573, 6.334)$ & $3.949$  & $3.893$  & $(3.031,5.106)$\\
				$\beta_{S}$  & $0.016$  & $0.016$  & $(-0.065, 0.097)$ & $0.006$  & $0.006$  & $(-0.051,   0.063)$\\				 
				$\beta_{H}$  & $-0.002$  & $-0.002$  & $(-0.023, 0.020)$  & $-0.010$  & $-0.010$  & $(-0.026, 0.006)$\\ 
				$\beta_{M}$  & $0.292$  & $0.284$  & $(-0.178, 0.793)$ & $0.299$  & $0.297$  & $(-0.041,   0.647)$\\
				\bottomrule
			\end{tabular}
			\par\end{centering}
	}
\end{table}

We ran the proposed PCG sampler with three over-dispersed initial values. 
Figure~\ref{real:convergence} shows the convergence characteristics of the sampler by using the $R^{1/2}$ diagnostic for the fixed-dimensional parameters and the fixed points of the varying coefficients \citep{gelm:rubi:92}. Because all $R^{1/2}$ statistics are below 1.1, we combine the second halves of three chains each with 150,000 iterations through a label switching algorithm. Then, after thinning every 50th sample,  our posterior inference is based on 4,500 mixed samples. 

 According to our posterior analysis, there are two main groups and a few minor groups. The group membership is determined by the posterior modes of the cluster labels. The interpretation of the analysis focuses on the two main groups. The first largest cluster, Group~1, accounts for 89.4\% of subjects, the second largest, Group~2, accounts for 7.2\%, and the remaining clusters account for 3.4\%. Some characteristics of the two major groups are summarized in Table~\ref{Characteristics}, demonstrating that these groups are made up of heterogeneous subjects. Specifically, Group~1 has a much lower proportion of females who tend to be more responsible for parenting than Group~2. In addition, Group~1 has a higher proportion of white-collar workers, civil servants, and university graduates with high job security than Group~2. Such difference in characteristics results in the different posterior estimates of varying-coefficient functions, as shown in Figure~\ref{univ:real_varying}.
	
	Figure~\ref{univ:real_varying} shows the posterior summaries of the varying-coefficient functions resulting from functional clustering. The first row of Figure~\ref{univ:real_varying} corresponds to the group-level varying-intercept functions. The intercept function of Group~1 is significantly positive and keeps increasing up to early 50s, implying that a posterior probability of being employed becomes higher as one tends to be old while holding all covariates constant. In contrast, Group~2 has a slightly positive but constant intercept function in all ages. The second row of Figure~\ref{univ:real_varying} shows the varying-coefficient function for having children below age 16 in the household. The 95\% pointwise posterior intervals for Group~1 includes 0, which implies that having children below age 16 in the household does not significantly affect the probability of employment. Unlike Group~1, the existence of young children in Group~2's household significantly decreases the probability of employment until his/her mid 40s. The probability is further decreased when the employee's age tends to be younger. 	This is due in part to the fact that Group~2 has a higher proportion of females than Group~1, and females were more responsible for child care in the late twentieth century.
	
	\begin{figure}[!t]
		\centering
		\psfrag{s1}[][][0.6]{$~~~~\alpha_{\cdot 0}^*(\cdot)$}
		\psfrag{s2}[][][0.6]{$~~~~\alpha_{\cdot K}^*(\cdot)$}
		
		\psfrag{j1}[][][0.6]{$~~~~\alpha_{0}(\cdot)$}
		\psfrag{j2}[][][0.6]{$~~~~\alpha_{K}(\cdot)$}
		
		\psfrag{a}[][][0.6]{Age}
		
		\includegraphics[width=0.55\textheight]{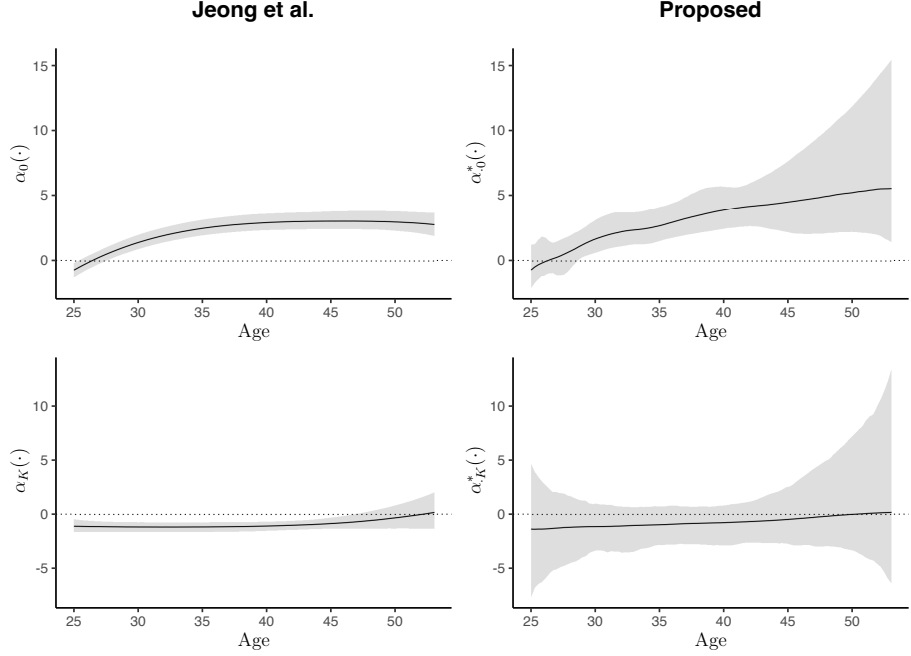}

		\caption
		{The first column shows the posterior summaries of the varying coefficients obtained by \cite{jeon:etal:17}, and the second column is with respect to the proposed method pooled by the weights corresponding to cluster assignments, with posterior medians (solid lines) and pointwise 95\% posterior intervals (gray areas).}
		\label{real:combined}
	\end{figure}
	
	Table~\ref{summary} shows the posterior summaries of fixed-dimensional parameters, where  ${\rm Var}(b_{i})=\psi$ represents the variance of a random effect, and $\beta_{H}$, $\beta_{S}$, and $\beta_M$ represent the coefficients of fixed effects, $H_{ij}$, $S_{ij}$, and $M_{ij}$, respectively. Based on the fact that the 95\% posterior intervals include zero, we decide that the fixed effects have no significant influence on the employment status when the varying effects of having young children in the household for heterogeneous subpopulations are accounted for in the model. When the heterogeneous subpopulation assumption is ignored, the results obtained by~\cite{jeon:etal:17} appear to be similar, but with narrower 95\% credible intervals. This is due to the fact that \cite{jeon:etal:17} does not fully account for the variability of heterogeneous subpopulations. 
	
Our proposed method identifies two major subpopulations with different characteristics, as shown in Table~\ref{Characteristics}, and these subpopulations show different age-varying effects of having young children in a household on working status, as illustrated in Figure~\ref{univ:real_varying}. When the heterogeneous subpopulation assumption is ignored, however, the single population model proposed by~\cite{jeon:etal:17} estimates the varying-coefficient functions applied to the entire population, as shown in the first column of Figure~\ref{real:combined}. In the presence of heterogeneous subpopulations, such an approach would fail to separate subpopulations with different characteristics, leading to erroneous conclusions. This is confirmed by producing the pooled varying-coefficient functions estimated by the proposed method, as shown in the second column of Figure~\ref{real:combined}. These findings demonstrate the proposed model's validity and utility in accounting for a heterogeneous population.

	

	\section{Discussion}
	\label{sec:disc}

In this paper, we propose a novel model-based functional clustering method for analyzing a heterogeneous series of binary data. Our proposed method models the varying effects of covariates on a series of binary responses as a function of an effect modifier, while accounting for heterogeneity among subjects using functional clustering and random effects. The proposed model estimates population-level fixed effects, cluster-level varying effects, and subject-level random effects. Even when the number of clusters is unknown, our proposed method accurately estimates cluster-specific varying coefficients with appropriate smoothness using a free-knot spline prior. We use the DP prior for functional clustering, which avoids specifying the exact number of clusters in advance. To perform posterior inference, we carefully develop a PCG sampler by specifying appropriate prior distributions and marginalizing some model components.

We suggest that there are several directions for future research, such as extending the clustering methodology to other generalized semi-parametric models using partitioning priors. Furthermore, a Gaussian process prior may be used instead of a free-knot spline for the functional clustering of varying coefficients because it may easily achieve the right scale for the base measure of the DP prior by employing a suitable covariance kernel.

	\appendix
	
	\section{Details of Algorithm~\ref{pcg.algorithm}}
	\label{sec:appen}
	
	In this section, we describe the details of Algorithm~\ref{pcg.algorithm}. Let $\mathcal{C}_{d}$ denote a set of clusters containing at least one subject and let $\mathcal{C}_{d}^{c}$ denote a set of clusters containing no subject, in the $d$th sampling iteration. For $d$th iteration:
	
	\begin{enumerate}[wide, labelwidth=!, labelindent=0pt, label = {\bf Step \arabic*}:,]
		\item Draw $\bga_{kl}^{*}$ from $p(\bga_{kl}^{*}|\bga_{-kl}^{*}, \bbe, \C, \V, \btau, \bPsi, \Li, \Y)$ that is a Bernoulli with success probability
		\begin{align*}
			\frac{f(\bga_{kl}^{*}=1,\bga_{-kl}^{*})}{f(\bga_{kl}^{*}=1,\bga_{-kl}^{*})+f(\bga_{kl}^{*}=0,\bga_{-kl}^{*})}, \quad k\in \mathcal{C}_{d},\quad l=1,\dots, p,
		\end{align*}
		where $\bga_{-kl}^{*}$ denotes all latent indicator variables except
		$\bga_{kl}^{*}$ in $\bga_{k}^{*}$, 
		\begin{align*}
			f(\bga_{kl}^{*},\bga_{-kl}^{*}) &~=~ B\left(|\bs{\gamma}_{kl}^{*}|+a, M_{l}+1-|\bs{\gamma}_{kl}^{*}|+b\right)\times\det\left(\tau_{k}\R_{k(\bga_{k}^{*})}^{-1}\Xi_{k(\bga_{k}^{*},\C,\bPsi)}+ \mf I_{|\bga_{k}^{*}|} \right)^{-1/2} \\
			&\qquad\times \exp\left\{\dfrac{1}{2}\xi_{k(\bga_{k}^{*},\C,\bPsi,\Li,\bbe)}^{\top}\left(\Xi_{k(\bga_{k}^{*},\C,\bPsi)}+\tau_{k}^{-1}\R_{k(\bga_{k}^{*})}\right)^{-1}
			\xi_{k(\bga_{k}^{*},\C,\bPsi,\Li,\bbe)}\right\}, \\
			\Xi_{k(\bga_{k}^{*},\C,\bPsi)}&~=~\sum_{i:C_{i}=k}\W_{i(\bga_{k}^{*})}^{\star\T}	\bT_i^{-1}\W_{i(\bga_{k}^{*})}^\star,\\	\xi_{k(\bga_{k}^{*},\C,\bPsi,\Li,\bbe)}&~=~\sum_{i:C_{i}=k}\W_{i(\bga_{k}^{*})}^{\star\T}\bT_i^{-1}\left(\Li_{i}-\X_{i}\bbe\right),
		\end{align*}
		and $\I_{|\bga_{k}^{*}|}$ and $\I_{n_{i}}$ are identity matrices whose sizes of each dimension are $|\bga_{k}^{*}|$ and $n_i$ respectively; and $\bT_i = \I_{n_{i}}+\Z_{i}\bPsi\Z_{i}^{\T}$.
		In the case of $k\in \mathcal{C}_{d}^{c}$, $\bga_{kl}^{*}$ is drawn from its prior distribution because $\Xi_{k(\bga_{k}^{*},\C,\bPsi)}$ and $\xi_{k(\bga_{k}^{*},\C,\bPsi,\Li,\bbe)}$ do not exist. 
		
		\item Draw $V_{k}$ from $p(V_k|\bga^{*}, \bbe, \C, \btau, \bPsi, \Li, \Y)$ that is a beta, i.e.,
		\begin{align*}
			V_{k}|(\bga^{*}, \bbe, \C, \btau, \bPsi, \Li, \Y) ~\sim~{\rm Beta}\left(1+m_{k},\nu+\sum_{h=k+1}^{K}m_{h}\right), \quad k=1, \dots, K-1,
		\end{align*}
		where $m_{k}=\sum_{i=1}^{N}I(C_{i}=k)$.
		
		\item Draw $\bphi_{(\bga_k^{*})}^{\star}$ from $p(\bphi_{(\bga_k^{*})}^{\star}|\bga^{*}, \bbe, \C, \V, \btau, \bPsi, \Li, \Y)$
		that is a multivariate normal distribution, i.e.,
		\begin{align*}
			&\bphi_{(\bga_{k}^{*})}^{\star}|(\bga^{*}, \bbe, \C, \V, \btau, \bPsi, \Li, \Y)\\
			& \quad \sim~{\rm N}_{|\bga_{k}^{*}|}\left(\left(\Xi_{k(\bga_{k}^{*},\C,\bPsi)}+\tau_{k}^{-1}\R_{k(\bga_{k}^{*})}\right)^{-1}\xi_{k(\bga_{k}^{*},\C,\bPsi,\Li,\bbe)}, \left(\Xi_{k(\bga_{k}^{*},\C,\bPsi)}+\tau_{k}^{-1}\R_{k(\bga_{k}^{*})}\right)^{-1}\right),\quad k\in \mathcal{C}_{d},
		\end{align*}
		and 
		\begin{align*}
			\bphi_{(\bga_{k}^{*})}^{\star}|(\bga^{*}, \bbe, \C, \V, \btau, \bPsi, \Li, \Y)~\sim~{\rm N}_{|\bga_{k}^{*}|}\left(\mathbf{0}, \tau_{k}\R_{k(\bga_{k}^{*})}^{-1}\right),\quad k\in \mathcal{C}_{d}^{c}.
		\end{align*}

		\item Draw $\bbe$ from $p(\bbe|\bphi_{(\bga^{*})}^{\star}, \bga^{*}, \C, \V, \btau, \bPsi, \Li, \Y)$ that is multivariate normal, i.e.,
		\begin{align*}
			& \bbe|(\bphi_{(\bga^{*})}^{\star}, \bga^{*}, \C, \V, \btau, \bPsi, \Li, \Y)~\sim~ {\rm N}_q\!\left( \boldsymbol\Delta^{-1}\sum_{i=1}^{N}\X_{i}^{\top}\bT_i^{-1}\left(\Li_{i} -\W_{i(\bga_{C_i}^{*})}\*\bphi_{(\bga_{C_i}^{*})}^{\star} \right), \boldsymbol\Delta^{-1} \right),
		\end{align*}
	where $\boldsymbol\Delta=\PP^{-1}+\sum_{i=1}^{N}\X_{i}^{\top}\bT_i^{-1}\X_{i}$.		
		
		\item Draw $\b_{i}$ from $p(\b_i|\bphi_{(\bga^{*})}^{\star}, \bga^{*}, \bbe, \C, \V, \btau, \bPsi, \Li, \Y)$
		that is multivariate normal, i.e.,
		\begin{align*}
			\b_{i}|(\bphi_{(\bga^{*})}^{\star}, \bga^{*}, \bbe, \C, \V, \btau, \bPsi, \Li_{i}, \Y)
			~\sim~{\rm N}_{r}\left(\U_{i(\bPsi,\bphi_{(\bgam_{C_i}^{*})}^{\star},\bgam^{*}_{C_i},\Li,\bbe)},\;\A_{(\bPsi)}\right),\quad i=1,\dots N,
		\end{align*}
		where
		\begin{align*}
			\U_{i(\bPsi,\bphi_{(\bgam_{C_i}^{*})}^{\star},\bgam^{*}_{C_i},\Li_{i},\bbe)}  &~=~ \bPsi\mf Z_{i}^{\top}(\mf I_{n_{i}}+\mf Z_{i}\bPsi\mf Z_{i}^{\top})^{-1}(\mf L_{i}-\mf W_{i(\bga_{C_{i}}^*)}^{\star}\bphi_{(\bga_{C_{i}}^*)}^{\star}-\X_{i}\bbe),\\
			\A_{(\bPsi)} &~=~\bPsi-\bPsi\mf Z_{i}^{\top}(\mf I_{n_{i}}+\mf Z_{i}\bPsi\mf Z_{i}^{\top})^{-1}\mf Z_{i}\bPsi.
		\end{align*}		
		
		\item Draw $\tau_{k}$ from $p(\tau_{k}|\bphi_{(\bga^{*})}^{\star}, \bga^{*}, \bbe, \b, \C, \V,  \bPsi, \Li, \Y)$
		that is an inverse gamma, i.e.,
		\begin{align*}
			&\tau_{k}|(\bphi_{(\bga^{*})}^{\star}, \bga^{*}, \bbe, \b, \C, \V, \bPsi, \Li, \Y)\\
			&\quad\sim~
				{\rm IG}\left(\dfrac{1+|\bs{\gamma}_{k}^{*}|}{2},\;\dfrac{N +\bphi_{(\bs{\gamma}_{k}^{*})}^{\star\top}\R_{k(\bga_{k}^{*})}\bphi_{(\bs{\gamma}_{k}^{*})}^{\star}}{2}\right),\quad k=1,\dots,K.
		\end{align*}
		
		\item Draw $\bPsi$ from  $p(\bPsi|\bphi_{(\bga^{*})}^{\star}, \bga^{*}, \bbe, \b, \C, \V,  \btau, \Li, \Y)$ that is an inverse Wishart, 
		\begin{align*}
			\bPsi \sim {\rm IW}\left(u+N, \D+\sum_{i=1}^{N}\b_{i}\b_{i}^{\top}\right).
		\end{align*}
		
		\item Draw $L_{ij}$ from $p(L_{ij}|\bphi_{(\bga^{*})}^{\star}, \bga^{*}, \bbe, \b, \C, \V,  \btau, \bPsi, \Y)$ that is truncated normal, i.e., 
		\begin{align*}
			L_{ij}|(\bphi_{(\bga^{*})}^{\star}, \bga^{*}, \bbe, \b, \C, \V, \btau, \bPsi, \Y)
			&\sim
			\begin{cases}
				{\rm TN}_{(-\infty,0]}(\boldsymbol{\mu}_{C_i}^{(j)}, 1) & \text{if \ensuremath{Y_{ij}=0}}\\
				{\rm TN}_{(0,\infty)}(\boldsymbol{\mu}_{C_i}^{(j)}, 1) & \text{if \ensuremath{Y_{ij}=1}}
			\end{cases},~ i=1,\dots, N,~ j=1,\dots, n_{i},
		\end{align*}
		where $\boldsymbol{\mu}_{C_i}^{(j)}$ 
		denotes the $j$th element of  $\boldsymbol{\mu}_{C_i} = \W_{i(\bga_{C_{i}}^{*})}^{\star}\bphi_{(\bga_{C_{i}}^{*})}^{\star}+\X_{i}\bbe+\Z_{i}\b_{i}$.
		
		\item Draw $C_{i}$ from $p(C_i|\bphi_{(\bga^{*})}^{\star}, \bga^{*}, \bbe, \b, \V, \btau, \bPsi, \Li, \Y)$ that has a discrete distribution with probabilities
		\begin{align*}
			&P(C_{i}=k|(\bphi_{(\bga^{*})}^{\star}, \bga^{*}, \bbe, \b, \V, \btau, \bPsi, \Li, \Y)~\propto~\dfrac{\pi_{k}(\V){\rm N}_{n_{i}}\left(\Li_{i};\boldsymbol{\mu}_{k},\:\mf I_{n_{i}}\right)}{\sum_{k=1}^{K}\pi_{k}(\V){\rm N}_{n_{i}}\left(\Li_{i};\boldsymbol{\mu}_{k},\:\mf I_{n_{i}}\right)},\quad k=1,\dots,K,
		\end{align*}
		where $\boldsymbol{\mu}_{k}=\W_{i(\bga_{k}^{*})}\*\bphi_{(\bga_{k}^{*})}^{\star}+\X_{i}\bbe+\Z_{i}\b_{i}$.
		
	\end{enumerate}

	

	\section{R package \texttt{fvcc}}
	\label{sec:appen_rpackage}
	We provide an R package called \texttt{fvcc} for the proposed model. The package can be installed with the \texttt{devtools} package available in CRAN as follows.
	\begin{verbatim}
		devtools::install_github("jwsohn612/fvcc")
		library(fvcc)
		help(fvcc)
	\end{verbatim}
	The main function \texttt{fvcc} contains a code script for reproducing the simulation results in Section~\ref{sec:sim}.
	

%
%
	
	\section*{Acknowledgements}
	S. Jeong's research was supported by the Yonsei University Research Fund of 2022-22-0097.
	T. Park's research was supported by Basic Science Research Program through the National Research Foundation of Korea(NRF) funded by the Ministry of Education (2017R1D1A1B03033536) and by the National Research Foundation of Korea(NRF) grant funded by the Korea government(MSIT) (2020R1A2C1A01005949).
	
	
	
	

	\bibnote{riph:etal:03}{[dataset]}	
	
	\bibliographystyle{elsarticle-num-names} 
	\bibliography{ref.bib}
	
\end{document}